\documentclass[aps,prd,twocolumn,showpacs]{revtex4} 

\usepackage{graphicx}

\font\FermiSmallfont=cmssq8 scaled 1200

\def\LANLppthead#1{
\null 
\begin{center}\vskip -1.0truein{\hbox to 7.5truein {
\hfill
\vbox to 1in {\vfill \FermiSmallfont
              \hbox{#1}
              \vfill}
}}\vskip-0.0truein\end{center}}

\begin{document}
\LANLppthead {LA-UR 06-2383}

\title{Light Element Signatures of Sterile Neutrinos and Cosmological Lepton Numbers}
\author{Christel J. Smith$^{1}$, George M.\ Fuller$^{1}$, Chad T. Kishimoto$^{1}$, Kevork N. Abazajian$^{2,3}$}
\affiliation{$^1$Department of Physics, University of California, San Diego, La Jolla, CA
92093-0319\\ $^2$Theoretical Division, MS B285, Los Alamos National Laboratory,
 Los Alamos, NM 87545 \\ $^3$Department of Physics, University of Maryland, College Park, MD
20742
}

\date{\today}

\begin{abstract}
We study primordial nucleosynthesis abundance yields for assumed ranges of cosmological lepton numbers, sterile neutrino mass-squared differences and active-sterile vacuum mixing angles. We fix the baryon-to-photon ratio at the value derived from the cosmic microwave background (CMB) data and then calculate the deviation of the $^2$H, $^4$He, and $^7$Li abundance yields from those expected in the zero lepton number(s), no-new-neutrino-physics case. We conclude that high precision ($< 5\%$ error) measurements of the primordial $^2$H abundance from, {\it e.g.}, QSO absorption line observations coupled with high precision  ($< 1\%$ error) baryon density measurements from the CMB could have the power to either: (1) reveal or rule out the existence of a light sterile neutrino if the sign of the cosmological lepton number is known; or (2) place strong constraints on lepton numbers, sterile neutrino mixing properties and resonance sweep physics. Similar conclusions would hold if the primordial $^4$He abundance could be determined to better than $10\%$.
\end{abstract}
\maketitle

\section{Introduction}
\label{sec-1}
Recent developments in observational cosmology may allow the primordial abundances of the light elements, including deuterium, to become novel probes of the mass and mixing properties of light sterile neutrinos.  The high precision inference of the baryon-to-photon ratio $\eta$ from the observed relative acoustic peak amplitudes in the cosmic microwave background (CMB) \cite{WMAP, WMAP1,3yrwmap} suggests a new way to employ Big Bang Nucleosynthesis (BBN) considerations. 

Historically, the comparison of the observationally-inferred light element abundances with calculated BBN abundance yields has been carried out with the intent of obtaining the baryon density. Indeed, on account of the near-exponential dependence of the deuterium yield on $\eta$ in BBN, the observations of isotope-shifted hydrogen absorption lines in Lyman limit systems along lines of sight to high redshift QSO's provide another, independent high precision measure of the baryon content of the universe \cite{Tytler,OMeara}. This value of $\eta$ currently is in good agreement, within errors, with the CMB-derived value \cite{burles,cyb03}. 

However, these two independent determinations of the baryon density depend on new neutrino physics in different ways. In particular, the BBN deuterium yield depends, albeit weakly, on the neutron-to-proton ratio and the expansion rate at the BBN epoch \cite {CF} and these quantities, in turn, can depend on the mass/mixing properties of sterile neutrinos. 

It has been shown recently that the $^4$He abundance yield in Big Bang Nucleosynthesis (BBN) can be dramatically sensitive to medium-enhanced active-sterile neutrino flavor transformation in the presence of a significant lepton number \cite{abfw,kfs}. This sensitivity comes about through (post-neutrino-decoupling) neutrino flavor mixing-induced alterations in the $\nu_e$ and $\bar\nu_e$ energy spectra. These alterations cause changes in the weak interaction rates governing the inter-conversion of neutrons and protons, and so ultimately  they cause changes in the neutron-to-proton ratio in BBN. We show here that active-sterile neutrino mixing likewise can induce modest changes in the deuterium and $^7$Li abundance yields.

This sets up a potentially new avenue for probing or constraining the active-sterile neutrino mixing parameter space: comparison of the value of $\eta$ derived from observationally-inferred deuterium on the one hand and the CMB-derived value on the other. Though the $^4$He BBN yield is far more sensitive to alterations of the neutron-to-proton ratio than is the $^2$H yield, at present the prospects for reliable and precise determination of the primordial deuterium abundance might be better than for helium.

The primordial helium abundance is likely between $23\%$ and $26\%$ by mass \cite{OS,Olive,IT}. It may be possible to do much better than this by adroit attention to issues of radiative transfer and compact blue galaxy morphology \cite{cyburt,Steigman,Steigman1}. However, as more QSO lines of sight become available, the statistics for deuterium abundance determinations in quasar absorption line systems will improve. Arguably, we may already know the primordial deuterium abundance at least as well as we know helium\cite{Tytler,OMeara,pettini}. In any case, it is worth exploring how much leverage deuterium measurements have in constraining the parameter space of sterile neutrino mass/mixing values and lepton number(s).

The LSND anomaly is being re-investigated in the mini-BooNE experiment \cite{LSND,miniB}. A positive signal in that experiment would indicate active neutrino coherent flavor transformation at a mass-squared scale significantly different from the atmospheric and solar neutrino mass-squared differences, $\delta m^2_{\rm atm} \approx 3\times{10}^{-3}\,{\rm eV}^2$ and $\delta m^2_{\odot} \approx 8\times{10}^{-5}\,{\rm eV}^2$, respectively. Given the $Z^0$-width limit, this would immediately imply the existence of a light sterile neutrino. 

If this light sterile neutrino (and its helicity-flipped partner, or \lq\lq sterile antineutrino\rq\rq ) were to completely thermalize in the early universe, there could be both an increased $^4$He yield, which is possibly unwelcome, and trouble with CMB- and large scale structure-derived bounds on the sterile neutrino rest mass closure contribution \cite{WMAP, WMAP1,abfw,Sloan,hannestad,murayama,aba, dib,dibad,barger,Kneller:2001cd,dod06}. Invocation of a significant net lepton number can suppress the production of sterile neutrinos in the epoch when neutrinos scatter frequently ({\it i.e.,} prior to weak decoupling), thereby easing these constraints \cite{FV,fv97,cirelli}. However, this lepton number will drive active-sterile resonant production of the sterile neutrino (or sterile antineutrino) after weak decoupling \cite{abfw}. Post-weak-decoupling resonant sterile neutrino production would leave the active neutrinos and the sterile neutrino with distorted, non-thermal energy spectra which can have a significant impact on the neutron-to-proton ratio and the $^4$He yield \cite{abfw}.

To investigate the effects of these spectral distortions and resonant transformation scenarios on the $^4$He, $^2$H and $^7$Li BBN yields, we follow the evolution of the neutrino distribution functions in various resonance sweep scenarios and self-consistently couple this with a calculation of the light element abundances performed with the full BBN nuclear reaction network code. (We employ a modified version of the Kawano/Wagoner code described in Ref.\ \cite{Smith:1992yy}.) In Section II we briefly outline the physics of active-sterile resonance sweep in the early universe. We describe our nucleosynthesis calculations in Section III. Results are given in Section IV. Discussion and conclusions are given in Section V.

\section{Resonant Active-Sterile Neutrino Flavor Transformation in the Early Universe}

Invocation of a significant lepton number as a dodge to full population of both helicity states of a sterile neutrino in the early universe \cite{FV,fv97,abfw,cirelli} will imply at least some resonant, medium-enhanced destruction of this lepton number and the concomitant production of sterile neutrinos \cite{abfw,cirelli}. In Ref.\ \cite{abfw} this general picture of post-weak-decoupling active-sterile resonance sweep in  the presence of a net lepton number was laid out in detail. The single channel active-sterile neutrino conversion problem posed in Ref.\ \cite{abfw} has recently been solved \cite{kfs}.

The weak decoupling epoch is where active neutrinos cease to scatter rapidly enough to exchange energy effectively with the background plasma. This takes place when the temperature is $T \approx 3\,{\rm MeV}$. Any neutrino energy spectral distortions which develop after this epoch will not be entirely erased by scattering and emission/absorption processes.

A pre-existing net lepton number in any of the neutrino flavors can drive medium-enhanced active-sterile neutrino flavor transformation in the early universe, both in the coherent regime after weak decoupling, and in the high temperature regime where de-coherence in the neutrino field becomes significant \cite{SF}. The lepton number residing in the neutrino sector associated with flavor $\alpha={\rm e},\mu,\tau$ is defined to be (in analogy to the baryon-to-photon ratio $\eta$)
\begin{equation}
L_{\nu_\alpha} \equiv {{n_{\nu_\alpha}-n_{\bar\nu_\alpha}}\over{n_\gamma}},
\label{lepton}
\end{equation}
where $n_{\nu_\alpha}$, $n_{\bar\nu_\alpha}$, and $n_\gamma$ are the neutrino, antineutrino, and photon number densities at some epoch. The potential lepton number corresponding to active neutrino flavor $\alpha$ is 
\begin{equation}
{\cal{L}}_\alpha \equiv 2 L_{\nu_\alpha} + \sum_{\beta \neq \alpha}{ L_{\nu_\beta}},
\label{potlepton}
\end{equation}
where also $\beta={\rm e},\mu,\tau$. (Note that neither $L_{\nu_\alpha}$ or ${\cal{L}}_\alpha$ are comoving invariants; we quote values of these assuming no dilution from $e^\pm$ annihilation, {\it i.e.,} at epoch $T \approx 3\,{\rm MeV}$.)

\begin{figure}
\includegraphics[width=2.5in,angle=270]{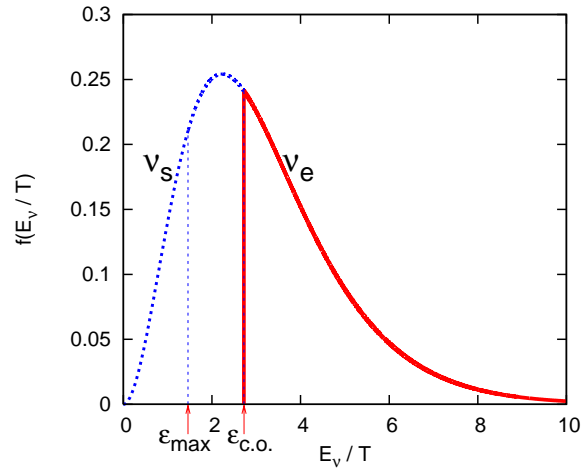}
\caption{Example active neutrino distribution function for a forced, adiabatic resonance sweep scenario.}
\label{figure1}
\end{figure}

\begin{figure}
\includegraphics[width=2.5in,angle=270]{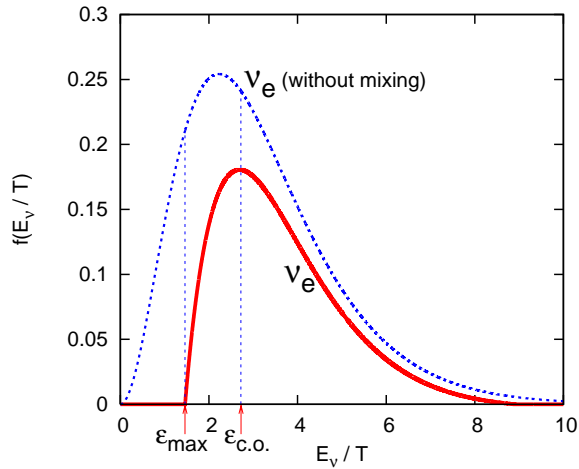}
\caption{Active neutrino distribution function obtained from the active-sterile transformation solution in Ref. \cite{kfs}.}
\label{figure2}
\end{figure}

It is convenient to cast the neutrino transformation problem in terms of the scaled-energy $\epsilon \equiv E_\nu/T$, instead of the neutrino energy $E_\nu$, because the former quantity is a co-moving invariant. A Mikheyev-Smirnov-Wolfenstein (MSW) \cite{MS,W} resonance, or neutrino mass level crossing, will occur between sterile and active neutrinos for electron neutrinos or antineutrinos for scaled-energy 
\begin{equation}
\epsilon_{\rm res} \approx {{\pi^2 \delta m^2 \cos2\theta}\over{4\sqrt{2}\, \zeta\left( 3\right)\,G_{\rm F} T^4}} \, \Bigg\vert {{1}\over{{\cal{L}}_e + \eta \left({{3}\over{2}} Y_e -{{1}\over{2}}  \right)}}\Bigg\vert,
\label{rese}
\end{equation}
and for mu and tau neutrinos or antineutrinos for scaled-energy
\begin{equation}
\epsilon_{\rm res} \approx {{\pi^2 \delta m^2 \cos2\theta}\over{4\sqrt{2}\, \zeta\left( 3\right)\,G_{\rm F} T^4}} \, \Bigg\vert{{1}\over{{\cal{L}}_{\mu,\tau} + \eta \left({{1}\over{2}} Y_e -{{1}\over{2}}  \right)}}\Bigg\vert,
\label{restau}
\end{equation}
where $\delta m^2$ is the mass-squared difference appropriate for the active-sterile mixing channel, $\zeta\left( 3\right) \approx 1.20206$ is the Riemann-Zeta function of argument $3$, $G_{\rm F}$ is the Fermi constant, and $Y_e = \left( n_{e^-}-n_{e^+}\right) n_\gamma^{-1} \eta^{-1}$ is the net electron number per baryon. Neutrinos transform at resonance if the terms inside the absolute value symbols in Eqs.~(\ref{rese}) and (\ref{restau}) are positive; antineutrinos transform if these terms are negative. In practice, for the rather large lepton numbers we employ, we can neglect the neutrino-electron scattering term (second term in the denominator within the absolute value) since $\eta \approx 6\times{10}^{-10}$ is so small.

The general picture of MSW resonance sweep ({\it i.e.}, how the resonance scaled-energy $\epsilon_{\rm res}$ depends on time/temperature) in the early universe is evident from Eqs.~(\ref{rese}) and (\ref{restau}). As the universe expands and the temperature drops, $\epsilon_{\rm res}$ will increase from zero.  If $\nu_{\alpha}$ neutrinos propagate through the MSW resonance coherently and adiabatically, they will be converted to sterile neutrinos with $100\%$ efficiency and, consequently, the lepton number will be depleted (${\cal{L}}_{\alpha}$ will decrease) \cite{SF,abfw}. Ref.\ \cite{abfw} showed that the resonance can sweep smoothly, continuously, and (most importantly) adiabatically only from $\epsilon_{\rm res} = 0$ to $\epsilon_{\rm res} = \epsilon_{\rm max}$. Here $\epsilon_{\rm max}$ is the value of the scaled energy where the product $\epsilon{\cal{L}}_{\alpha}$ is a maximum, $\epsilon^3_{\rm max} \approx  2\zeta\left( 3\right)\, \left(e^{\epsilon_{\rm max} - \eta_{\nu_\alpha}} + 1\right){\cal{L}}_{\alpha}\left(\epsilon_{\rm max}\right)$ \cite{abfw}. Ref.\ \cite{abfw} showed that the MSW resonance cannot sweep smoothly and adiabatically beyond this point, $\epsilon_{\rm res} = \epsilon_{\rm max}$.

If we force the resonance to continue to sweep adiabatically and continuously past $\epsilon_{\rm max}$, completely converting $\nu_{\alpha}$ neutrinos in the portion of the $\nu_{\alpha}$ distribution with scaled energy $E_{\nu}$/T $\leq \epsilon_{\rm res}$, we would completely deplete the lepton number (${\cal{L}}_{\alpha} = 0$) when the resonance reaches $\epsilon_{\rm res} = \epsilon_{\rm c.o.}$ (where "c.o." stands for cut-off). As an example, the initial and final distribution functions for a Fermi-Dirac $\nu_{e}$ energy spectrum with degeneracy parameter (chemical potential divided by temperature) $\eta_{\nu_{e}} = 0.05$ and potential lepton number ${\cal{L}} = 0.368$ for this forced adiabatic sweep scenario is shown in Figure \ref{figure1}. The resulting active neutrino spectrum in this case would have a low energy ``cut". That is, the distribution function would be zero for $E_{\nu}$/T $\leq \epsilon_{\rm c.o.}$.  The sterile neutrino produced in this scenario $\nu_{\alpha}\rightarrow\nu_{\rm s}$ would have a distribution function identical to the original $\nu_{e}$-spectrum for $E_{\nu}$/T $\leq  \epsilon_{\rm c.o.}$ but zero for larger values of scaled energy.

Recently the active-sterile resonance sweep problem for $\epsilon_{\rm res} > \epsilon_{\rm max}$ has been solved \cite{kfs}. The resonance does sweep continuously past $\epsilon_{\rm max}$ to (near) lepton number depletion, but it does so non-adiabatically. That is, for $\epsilon_{\rm res} > \epsilon_{\rm max}$, $\nu_{\alpha}\rightarrow\nu_{\rm s}$ is not $100\%$ efficient. The net result is that the resonance must sweep to higher energy to significantly deplete the lepton number. The resonance will sweep adiabatically as before to $\epsilon_{\rm max}$, but there will be non-adiabatic, incomplete conversion in the $\nu_{\alpha}\rightarrow\nu_{\rm s}$ channel as $\epsilon_{\rm res}$ sweeps to higher scaled energy, and then a return to complete, adiabatic conversion at large values of $\epsilon_{\rm res}$. This scenario is depicted in Figure \ref{figure2} for the case of $\nu_{e}$'s with an initial Fermi-Dirac spectrum and lepton numbers $L_{\nu_{e}} = 0.0343$, and $L_{\nu_{\mu}} = L_{\nu_{\tau}} = 0.15$. 

A distorted, non-thermal $\nu_{e}$ (or $\overline{\nu}_{e}$) spectrum will change the neutron-to-proton ratio in BBN and hence, the light element abundance yields over the case with thermal, Fermi-Dirac energy spectra \cite{abfw}. This is because the $\nu_{e}$ and/or $\overline{\nu}_{e}$ energy spectra determine the rates of the neutron and proton inter-conversion processes,
\begin{eqnarray}
& \nu_e+n\rightleftharpoons p+e^-, \label{nuen} \\
& \bar\nu_e+p\rightleftharpoons n+e^+, \label{nuebarp} \\
& n \rightleftharpoons p+e^-+\bar\nu_e. \label{ndecay}
\end{eqnarray}
For a given initial potential lepton number, removing neutrino population at higher scaled energy in the spectrum results in a larger effect on the neutron-to-proton ratio. This is because of the significant neutrino energy dependence in the cross sections and dsitribution functions which factor into the rates of the processes in Eqs.\ (\ref{nuen}), (\ref{nuebarp}), and (\ref{ndecay}).  However, another factor in this behavior is that more neutrino population lies in the portion of the $\nu_{e}$ or $\overline{\nu}_{e}$ spectrum for $E_{\nu}$/T $\geq  \epsilon_{\rm c.o.}$ for the values of potential lepton number of most interest here.

\begin{figure}
\includegraphics[width=2.5in,angle=270]{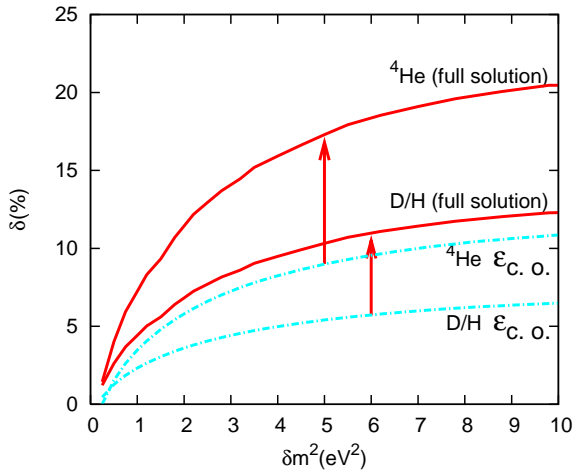}
\caption{Percent change in $^2$H and $^4$He yield from standard BBN calculated as a function of $\delta m^2$ for the two active-sterile mixing cases. The two cases are forced adiabatic resonance sweep to $\epsilon_{\rm c.o.}$ (Figure \ref{figure1}) and the full solution (Figure \ref{figure2}).}
\label{figure3}
\end{figure}

For a given initial potential lepton number, the actual (full solution) resonance sweep scenario will give bigger BBN alteration effects than will the artificial forced continuous, adiabatic sweep model \cite {kfs}. This is shown in Figure \ref{figure3}, where we plot the fractional change (in percent) in the helium and deuterium abundance yields from their standard BBN values for the CMB-determined baryon density as a function of the mass-squared difference $\delta m^2$, characterizing the $\nu_{e}\rightarrow\nu_{\rm s}$ oscillation channel. We give this for both the adiabatic sweep to $\epsilon_{\rm c.o.}$ scenario and the full resonance sweep solution.  In these calculations we have taken the initial lepton numbers to be $L_{\nu_{e}} = 0.0343$ and $L_{\nu_{\mu}} = L_{\nu_{\tau}} = 0.15$.

In reality we could expect active-active neutrino conversion simultaneous with active-sterile transformation. This could leave complicated distortion features in the energy spectra of all neutrino species \cite{abfw}. It is the spectral distortions of $\nu_{e}$ and $\overline{\nu}_{e}$ neutrinos which are most important for BBN. Active-active neutrino mixing could tend to partially fill in the $E_{\nu}$/T $<  \epsilon_{\rm max}$ portion of the $\nu_{e}$ or $\overline{\nu}_{e}$ spectrum, though this could be offset by continued sweep to even higher $E_{\nu}$/T. In any case, to be conservative in our BBN abundance yield estimates we will in what follows assume a smooth, adiabatic resonance sweep to $\epsilon_{\rm c.o.}$ for a given potential lepton number.  Therefore, our calculated abundance yield changes for given lepton numbers and sterile neutrino mass/mixing data will be (usually) underestimates.

\section{Primordial Nucleosynthesis Calculations with Neutrino Spectral Distortions}

In general, active-sterile resonance sweep will go on simultaneously with the charged current weak interactions that set the neutron-to-proton ratio ($n/p$), as well as the strong and electromagnetic nuclear reactions associated with BBN. We will have distorted neutrino $\nu_e$ and $\bar\nu_e$ energy spectra with distortions that change in time as the active-sterile resonance sweeps and active-active neutrino flavor transformation proceeds. These new features necessitate handling the weak interactions differently than in the standard BBN case.

The standard BBN code, originally written by R. Wagoner \cite{wfh} and later revised by L. Kawano \cite{kawano,kawano1}, calculates the processes in Eqs.\ (\ref{nuen}), (\ref{nuebarp}), and (\ref{ndecay}) by adding the three $n\rightarrow p$ rates; likewise, the three $p\rightarrow n$ rates:
\begin{equation}
\lambda_{n} = \lambda_{\nu_e+n\rightarrow p+e^-} + \lambda_{n+e^+\rightarrow p+\bar\nu_e} + \lambda_{n\rightarrow p+e^-+\bar\nu_e}
\label{n-rates}
\end {equation}
\begin{equation}
\lambda_p = \lambda_{p+e^-\rightarrow \nu_e+n} + \lambda_{\bar\nu_e+p\rightarrow n+e^+} + \lambda_{p+e^-+\bar\nu_e\rightarrow n}.
\label{p-rates}
\end{equation}
In the standard calculations, the integrands in $\lambda_n$ and $\lambda_p$ are manipulated and condensed into a shorter, two-part integral to save computation time. This requires the neutrino and antineutrino distribution functions (as well as the electron and positron) be of Fermi-Dirac form. If the neutrino degeneracy parameters are zero, the code calculates $\lambda_n$ and $\lambda_p$ with a series approximation to further cut down computation time. This can lead to an erroneous $\approx 0.5\%$ increase in the neutron-to-proton ratio \cite{kawano,kawano1}. 

\begin{figure}
\includegraphics[width=2.5in,angle=270]{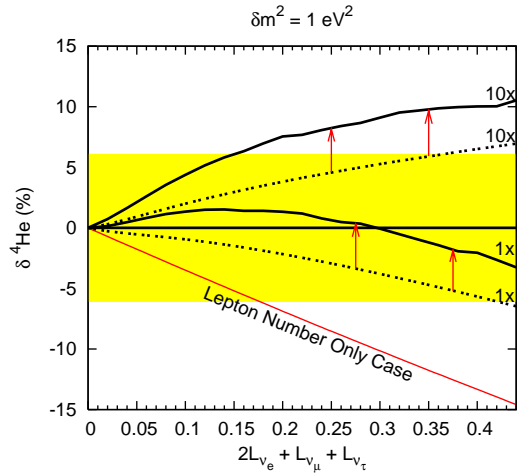}
\caption{Percent change in primordial $^4$He yield relative to the standard BBN zero lepton number, no sterile neutrino model is shown as a function of potential lepton number $\mathcal{L}_{e} = 2 L_{\nu_{e}} + L_{\nu_{\mu}} + L_{\nu_{\tau}}$. In all cases the baryon-to-photon ratio is fixed at the CMB-derived value $\eta = 6.11 \times 10^{-10}$. The central horizontal line is the standard BBN case. The light solid line is the case with lepton numbers but with no active-sterile neutrino mixing. In this case we set all individual lepton numbers to be equal. The lower dashed line is the case with active-sterile neutrino mixing in the forced, adiabatic resonance sweep scenario with lepton number distribution factor $= (L_{\nu_{\mu}} + L_{\nu_{\tau}})/{2 L_{\nu_{e}}} = 1$. The lower heavy solid line is the same case but in the full resonance sweep scenario. The upper dashed line similarly corresponds to the adiabatic resonance sweep scenario but with lepton number distribution factor 10, while the upper solid line is the same case in the full resonance sweep scenario.  All active-sterile mixing cases here have $\delta m^2 = 1\,{\rm eV}^2$. The light horizontal band gives the $23\% - 26\%$ range for $Y_p$ allowed by observational bounds.}
\label{figure4}
\end{figure}

\begin{figure}
\includegraphics[width=2.5in,angle=270]{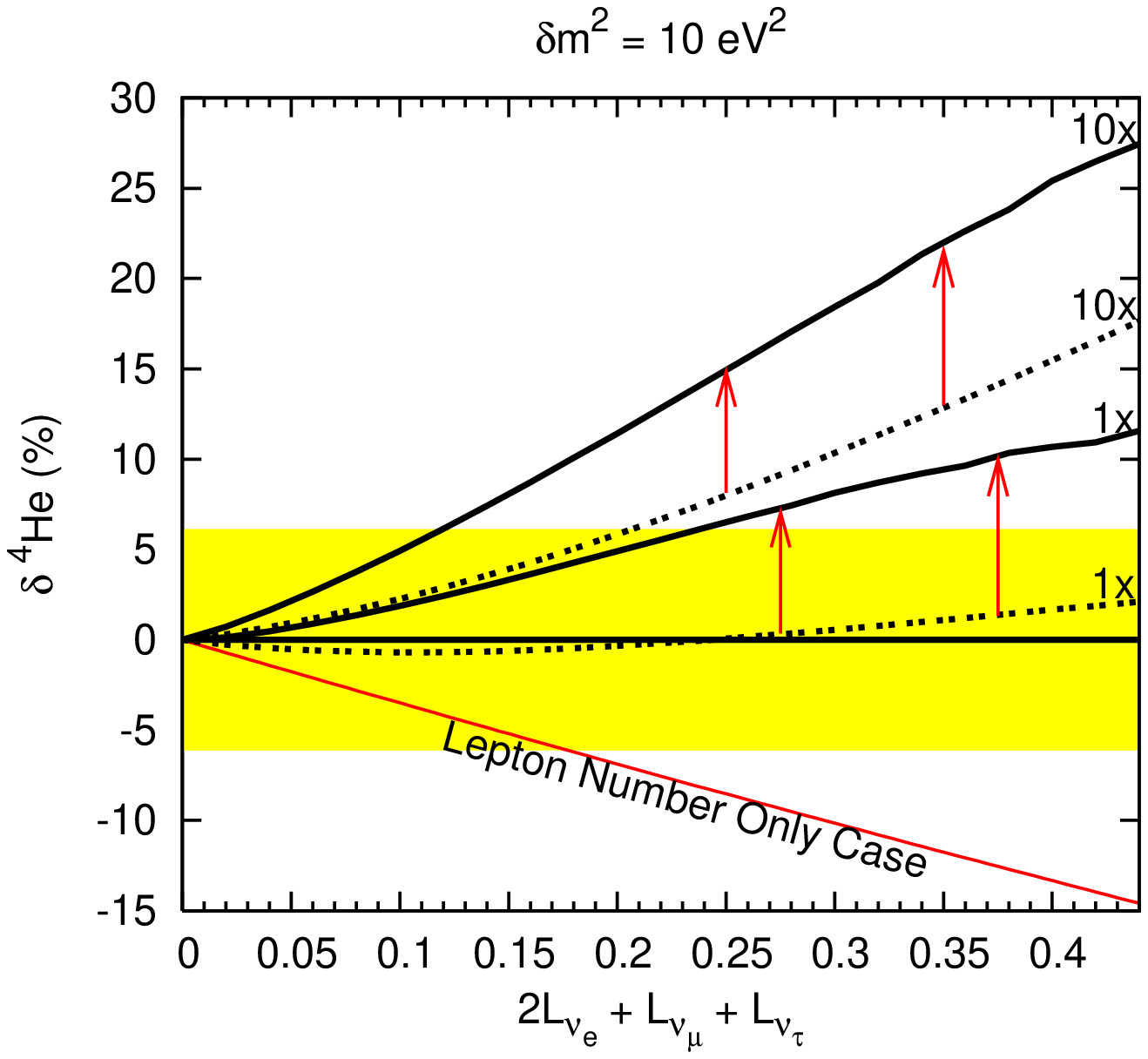}
\caption{Same as for Fig.\ \ref{figure4} but now with $\delta m^2 = 10\,{\rm eV}^2$.}
\label{figure5}
\end{figure}

\begin{figure}
\includegraphics[width=2.5in,angle=270]{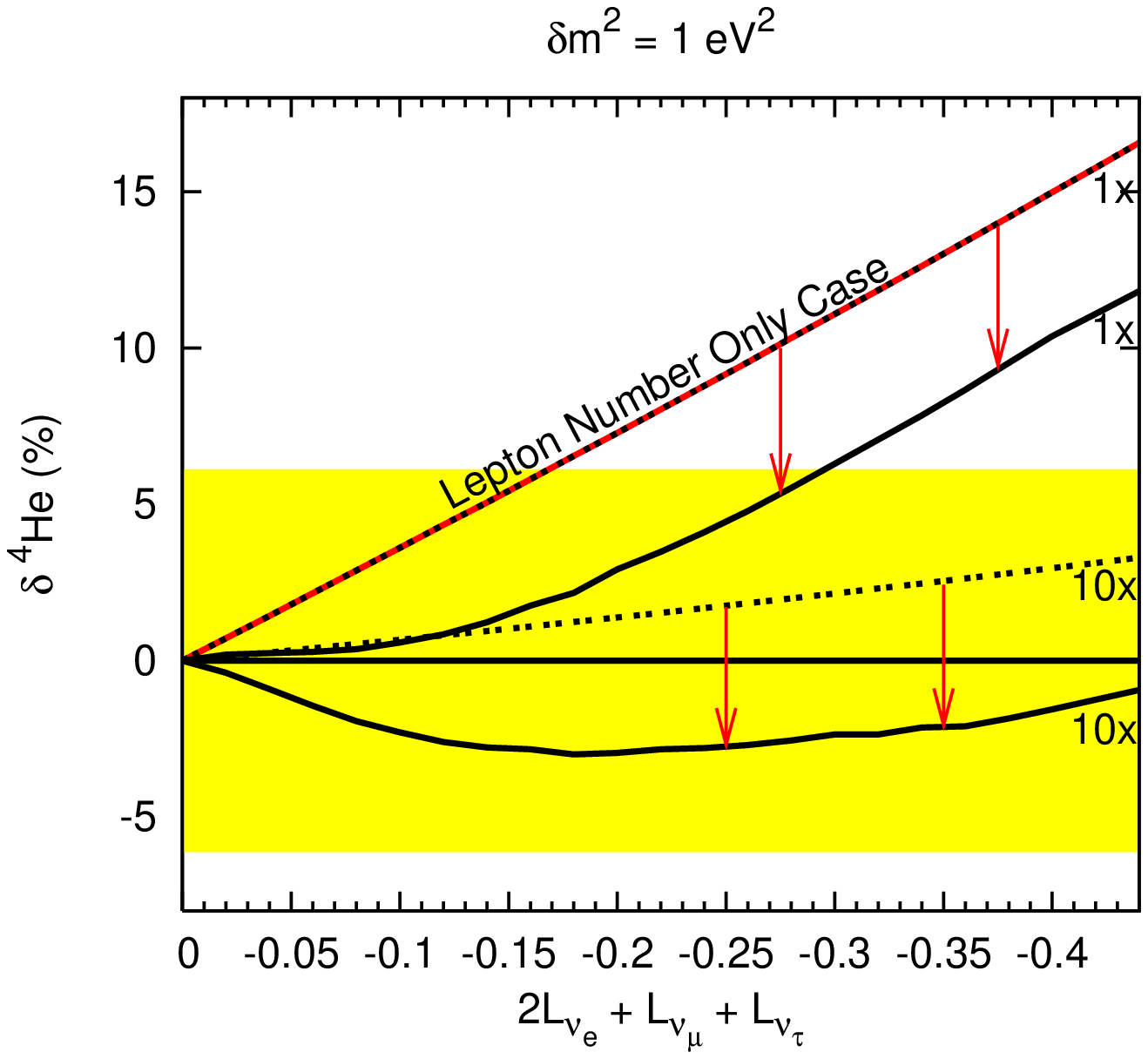}
\caption{Same as for Fig.\ \ref{figure4} but now with negative lepton numbers.}
\label{figure6}
\end{figure}

\begin{figure}
\includegraphics[width=2.5in,angle=270]{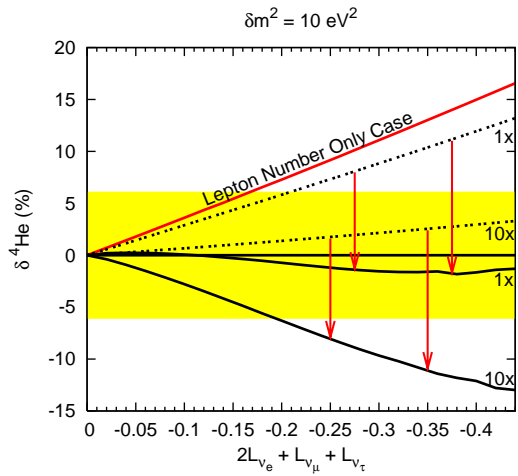}
\caption{Same as for Fig.\ \ref{figure5} but now with negative lepton numbers.}
\label{figure7}
\end{figure}

In order to implement time-dependent neutrino and antineutrino spectral distortions, we rewrote the BBN code to calculate each weak rate in Eqs.\ (\ref{nuen}), (\ref{nuebarp}), and (\ref{ndecay}) independently, with no series approximations, and changed the integration variable to neutrino/antineutrino  energy instead of electron/positron energy.
We then modularized the weak rate calculations so that any neutrino or antineutrino distribution function could be entered. Likewise, our modifications allow us to implement any desired time dependence in these distribution functions and they also allow us to calculate consistently the energy density in (and spectra of) any sterile neutrinos which are produced. These modifications are implemented in four modules. One module contains the matrix elements and phase space integrands for the weak rates. The phase space integrands call a second module which contains the $\nu_e$ and $\bar\nu_e$ distribution functions. The third module defines the limits of integration. Finally, the fourth module calculates the resonance energy for active-sterile transformation and calls the integrator, which in turn calls the other modules for the integrands and integration limits.  In addition, the expansion rate of the universe at any temperature is calculated self-consistently with all active and sterile neutrino distribution functions.

In all of our BBN calculations, we set the baryon-to-photon ratio to $\eta = 6.11\times 10^{-10}$. This corresponds to the central value of the CMB acoustic peak amplitude-determined WMAP Three Year Mean result, $\eta = (6.11\pm .22) \times 10^{-10}$, which, in turn, corresponds to a baryon rest mass closure fraction $\Omega_{\rm b}h^2 = 0.0223\pm0.0008$ \cite{3yrwmap}, where $h$ is the Hubble parameter in units of 100 km s$^{-1}$ Mpc$^{-1}$. 

In our calculations, for illustrative purposes, we adopt neutron lifetime $\tau_n = 887.8\,{\rm s}$.  The current world average \cite{pdg} for this lifetime is $\tau_n = 885.7 \pm 0.8\,{\rm s}$, but a recent measurement \cite{serebrov} suggests it could be as small as $\tau_n = 878.5 \pm 0.7 \pm 0.3\,{\rm s}$.  Our adopted $\tau_n$ is larger than all of these and this has the effect of making our distorted-neutrino-spectrum calculations {\it underestimates} of the actual change in the $n / p$-ratio and, hence, nucleosynthesis yield deviations.  Though these differences are small, it must be kept in mind that $\tau_n$ remains uncertain to a degree.

With these choices of $\eta$ and $\tau_n$, our modified version of the BBN code calculates the $^4$He mass fraction to be $Y_p = 0.2429$ and the deuterium abundance relative to hydrogen to be D/H = $2.543\times 10^{-5}$. Although the current uncertainty in the WMAP-derived baryon density is relatively small ($\sim \pm 3.6 \%)$, it nevertheless translates into a $\sim \pm 5.5\%$ uncertainty in the predicted primordial value of D/H. This is because the BBN deuterium yield is a very sensitive function of $\eta$. As we discuss below, the error in D/H stemming from the current error in the CMB-determined $\eta$ precludes using the observationally-determined deuterium abundance to constrain the sterile neutrino physics discussed in this paper. However, the higher precision determinations of $\eta$ in the projected Four Year WMAP results lead to an uncertainty of $\pm 0.00047$ in $\Omega_{\rm b}h^2$, corresponding to $\pm 0.117$ in $\eta$, while the forthcoming Planck mission forecasts $\pm 0.00017$ in $\Omega_{\rm b}h^2$, or $\pm 0.045$ in $\eta$ \cite{bond}. These more precise determinations of $\eta$ will translate into commensurately better precision in the calculated D/H values. As we discuss below, these could allow for new neutrino physics and/or lepton number constraints.

Corrections to the code, such as time-step corrections and Coulomb and radiative corrections, have been discussed extensively ({\it e.g.}, Ref.\ \cite{tandl}; Ref.\ \cite{kandk}).  Most corrections are a small additive factor to the final helium abundance $Y_p$ and are functions of the chosen time-step, $\eta$, and $\tau_n$.  Since this work uses set values for these parameters and presents the results in terms of percent change, the additive corrections do not contribute to the final results. 

It is beyond the scope of this work to present precision element abundance predictions. Our goal here is to illustrate the global trends in element production resulting from adding in the active-sterile transformation physics.  Eventually corrections, such as the coulomb correction, should be calculated autonomously in the weak rates in order to give the $<1\%$ accuracy desired.

\section{BBN Abundance Yields with Lepton Numbers and Sterile Neutrinos}

\begin{figure}
\includegraphics[width=2.5in,angle=270]{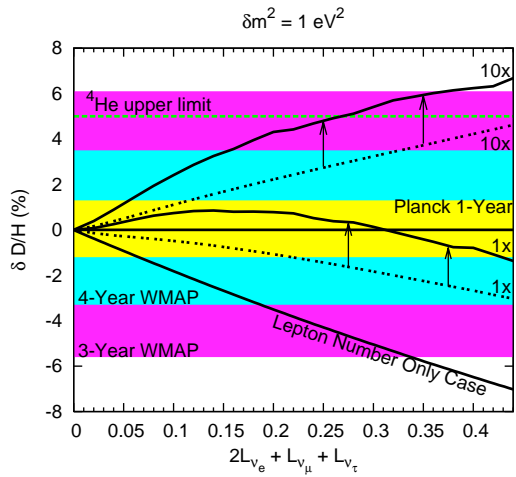}
\caption{Same as Fig.\ \ref{figure4} but now for the percent change in D/H, the deuterium abundance relative to hydrogen. Here the outer horizontal band gives the standard BBN range in D/H corresponding to the uncertainty in baryon-to-photon ratio $\eta$ for the Three Year WMAP data, while the middle and inner bands give the D/H range for the uncertainty in $\eta$ for the Four Year WMAP and the projected one year Planck result, respectively. The horizontal dashed line shows where neutrino spectral distortion plus lepton number will give a $^4$He yield exceeding $26\%$. Here $\delta m^2 = 1\,{\rm eV}^2$.}
\label{figure8}
\end{figure}

\begin{figure}
\includegraphics[width=2.5in,angle=270]{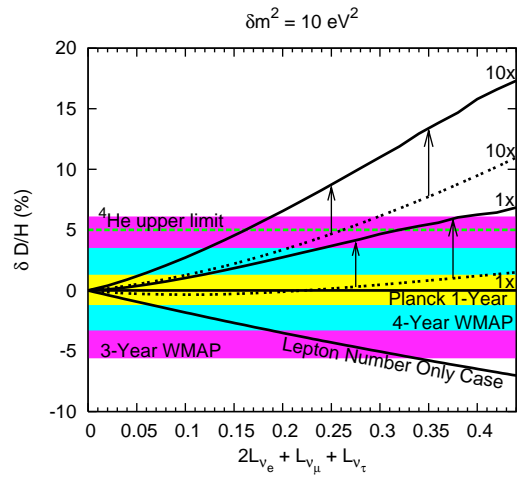}
\caption{Same as for Fig.\ \ref{figure8} but now with $\delta m^2 = 10\,{\rm eV}^2$.}
\label{figure9}
\end{figure}

Here we describe the results of our calculations of light element primordial nucleosynthesis in the presence of significant lepton numbers and active-sterile neutrino flavor mixing. The properties of light sterile neutrinos and the lepton numbers of the universe could be related \cite{SF,FV}, but here we shall vary them independently to gauge effects on BBN abundance yields. We therefore have five quantities to vary.

The first of these parameters is the rest mass of the sterile neutrino $m_{s}$ or, equivalently, the mass-squared difference $\delta m^{2} \approx m_{s}^{2}$ characteristic of active-sterile neutrino flavor mixing. The second quantity is the effective $2 \times 2$ vacuum mixing angle $\theta$ characterizing the unitary transformation between an active neutrino, which we will take to be electron flavor $\vert \nu_{e} \rangle$, and a sterile state $\vert \nu_{s} \rangle$ and mass/energy eigenstates $\vert \nu_{1} \rangle$ and $\vert \nu_{2} \rangle$,
\begin{eqnarray}
\vert \nu_{e} \rangle & = & \cos \theta \vert \nu_{1} \rangle + \sin \theta \vert \nu_{2} \rangle, \nonumber \\
\vert \nu_{s} \rangle & = & -\sin \theta \vert \nu_{1} \rangle + \cos \theta \vert \nu_{2} \rangle,
\end{eqnarray}
with corresponding rest-mass eigenvalues $m_{1}$ and $m_{2}$, respectively, such that $\delta m^{2} \equiv \vert m_{2}^{2} - m_{1}^{2} \vert$. Here we set $\sin^2 2 \theta = 10^{-3}$ to conform with the LSND results. Because the expansion rate of the universe is so slow at the epoch of medium-enhanced coherent MSW sterile neutrino production, flavor evolution is likely adiabatic, at least for scaled resonance energy $\epsilon_{\rm res} \leq \epsilon_{\rm max}$ \cite{abfw,kfs}. As a consequence, our nucleosynthesis results will change little with variation in $\theta$ so long as sin$^22\theta > 10^{-5}$. 

As outlined above and in Ref.\ \cite{abfw}, instead of treating the complete $4 \times 4$ mass/mixing matrix with its many unknown mixing parameters, we shall consider the effective $2 \times 2$ conversion channels $\nu_{e} \rightleftharpoons \nu_{s}$ or  $\bar\nu_{e} \rightleftharpoons \bar\nu_{s}$ and adopt two different resonance sweep schemes in an attempt to bracket the BBN effects of the active-active plus active-sterile mixing-induced spectral distortions. To follow resonance sweep we use: (1) continuous, adiabatic sweep to lepton depletion at $\epsilon_{\rm c.o.}$, and (2) the full solution of Ref. \cite{kfs}. The latter resonance sweep scheme gives the most dramatic alterations in $\nu_{e}$ or $\bar\nu_{e}$ energy distribution for a given lepton number, but in reality active-active mixing $\nu_{e} \rightleftharpoons \nu_{\mu,\tau}$ or $\bar\nu_{e} \rightleftharpoons \bar\nu_{\mu,\tau}$, as well as $\nu_{\mu,\tau} \rightleftharpoons \nu_{s}$ or $\bar\nu_{\mu,\tau} \rightleftharpoons \bar\nu_{s}$, will likely fill in some of the spectral deficits in $\nu_{e}$ or $\bar\nu_{e}$, as will post-decoupling inelastic neutrino and antineutrino scattering. By contrast, the continuous, adiabatic sweep to $\epsilon_{\rm c.o.}$ scenario gives conservative underestimates of BBN effects \cite{abfw}.

The remaining three parameters in our BBN calculations are the actual lepton numbers. For $\alpha = e, \mu, \tau$ we have
\begin{equation}
L_{\nu_{\alpha}} = \left( \frac{\pi^2}{12 \zeta (3)} \right) {\left(\frac{T_{\nu}}{T_{\gamma}}\right)}^3 \left[\eta_{\nu_{\alpha}} + \eta_{\nu_{\alpha}}^{3}/\pi^{2}\right],
\end{equation}
where $\zeta (3) \approx 1.20206$, $T_{\nu}$ and $T_{\gamma}$ are the neutrino and plasma temperature, respectively, and the ratio of neutrino chemical potential to neutrino temperature is the neutrino degeneracy parameter $\eta_{\nu_{\alpha}}$. While $\eta_{\nu_{\alpha}}$ is a co-moving invariant, $L_{\nu_{\alpha}}$ is not because the ratio $T_{\nu}/T_{\gamma}$ varies as temperature drops and the entropy initially in the seas of electrons and positrons is transferred to photons. (After all $e^{\pm}$-pairs have disappeared, $(T_{\nu}/T_{\gamma})^{3} = 4/11$.) The lepton numbers given in our figures are for $T_{\nu}/T_{\gamma}=1$. We assume that all neutrinos and antineutrinos initially have Fermi-Dirac equilibrium energy spectra ({\it e.g.}, the heavy dashed line in Fig.\ \ref{figure2}), so that $\eta_{\bar\nu_{\alpha}}=-\eta_{\nu_{\alpha}}$.

\begin{figure}
\includegraphics[width=2.5in,angle=270]{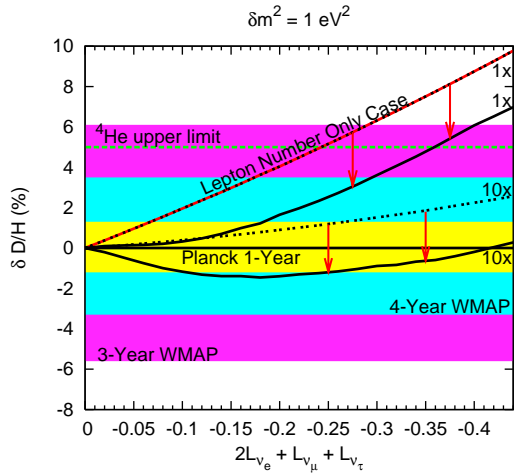}
\caption{Same as for Fig.\ \ref{figure8} but now with negative lepton numbers.}
\label{figure10}
\end{figure}

\begin{figure}
\includegraphics[width=2.5in,angle=270]{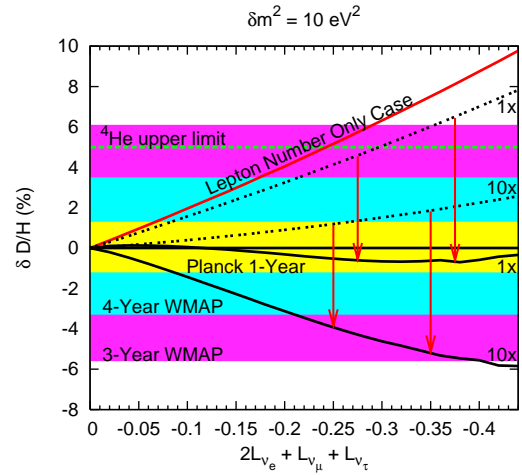}
\caption{Same as for Fig. \ref{figure9} but now with negative lepton numbers.}
\label{figure11}
\end{figure}

Active-active $3 \times 3$ neutrino mixing with the solar and atmospheric mass-squared differences has been shown to ``even up" the lepton numbers of each neutrino flavor to within a factor of $\sim$10 \cite{abb,wong,dolgov,smf}. This also may be true in the $4 \times 4$ mixing case. Since we consider $\nu_{e} \rightleftharpoons \nu_{s}$ or $\bar\nu_{e} \rightleftharpoons \bar\nu_{s}$, the relevant potential lepton number is $\mathcal{L}_{e} = 2 L_{\nu_{e}} + L_{\nu_{\mu}} + L_{\nu_{\tau}}$. The lepton number distribution factor 
\begin{equation}
{\rm factor} = \frac{L_{\nu_{\mu}} + L_{\nu_{\tau}}}{2 L_{\nu_{e}}}
\label{factor}
\end{equation}
is an important determinant of spectral distortion. We consider values of this factor between 1 and 10. The bigger the value of this factor, the larger will be the spectral distortion for a given value of $\eta_{\nu_{e}}$ or $\eta_{\bar\nu_{e}}$ \cite{abfw}. This is because the larger the potential lepton number, the larger will be, {\it e.g.}, $\epsilon_{\rm max}$ and $\epsilon_{\rm c.o.}$, quantities which set the scale for spectral distortion.

A positive $\nu_{e}$ degeneracy parameter $\eta_{\nu_{e}}$ reflects an excess of $\nu_{e}$'s over $\bar\nu_{e}$'s, which will have the effect of driving the reaction in Eq.\ (\ref{nuen}) to the right, thereby lowering $n/p$ and the $^{4}$He yield. However, $\nu_{e} \rightarrow \nu_{s}$ flavor conversion can effectively distort the $\nu_{e}$ spectrum so as to lower the overall $\nu_{e}$ number density and thereby shift the reaction in Eq.\ (\ref{nuen}) back to the left, in effect counteracting the $\nu_{e}$ degeneracy and, possibly, leading to an {\it increase} in the $^4$He yield over the standard BBN, $\eta_{\nu_{e}} =0$ result. This argument can be reproduced in analogous fashion for a $\bar\nu_{e}$ excess ($\eta_{\nu_{e}} < 0$) and $\bar\nu_{e} \rightleftharpoons \bar\nu_{s}$ flavor conversion.

Figures \ref{figure4} and \ref{figure5} show the percent change in the $^4$He primordial nucleosynthesis yield relative to the standard BBN model with no lepton numbers and no new neutrino physics as a function of (positive) potential lepton number. In all cases the baryon-to-photon ratio is set to the central CMB-derived value $\eta = 6.11 \times 10^{-10}$. With lepton numbers alone, but without sterile neutrino mixing and spectral distortion, the neutron-to-proton ratio is suppressed and the $^4$He yield is decreased relative to the standard model. This trend is weakened or even completely reversed when $\nu_e \rightleftharpoons \nu_s$ generated spectral distortion is included in the calculations.  In general, the spectral distortions generated in the full resonance sweep scenario produce bigger increases in abundance yield over the lepton-number-only case than does the forced adiabatic resonance sweep scenario.  As discussed above, this stems from the tendency in the full resonance sweep mechanism to deplete $\nu_e$ population at higher energy in the distribution function. From Figures \ref{figure4} and \ref{figure5} it is clear that the existence of a sterile neutrino could alter significantly the relationship between predicted $^4$He abundance and lepton numbers. It is also clear from these figures that improvement in the precision of the observationally-inferred value of $Y_p$ could make for stringent new constraints on the sterile neutrino parameters and lepton numbers. Even with the (likely overly) generous range of $23\% - 26\%$ for $Y_p$ we see that larger sterile neutrino masses together with larger positive lepton numbers tend to produce too much $^4$He.

Figures \ref{figure6} and \ref{figure7} likewise show the deviation in the $^4$He yield from the standard model value but now for negative lepton numbers. These figures are not simply mirror images of Figures \ref{figure4} and \ref{figure5}. This is because negative lepton numbers will produce distortions in the $\bar\nu_e$ energy spectrum. There is a threshold in the $\bar\nu_e$ capture process in Eq.\ (\ref{nuebarp}), while there is no threshold in the $\nu_e$ capture process in Eq.\ (\ref{nuen}). The result is that we must have a distortion extending to higher energy in the $\bar\nu_e$ distribution than in the $\nu_e$ distribution to produce the same magnitude change in $^4$He yield. This trend is obvious in Fig.\ \ref{figure4} where there is no discernible difference between the case with lepton number alone and the case with $\bar\nu_e \rightleftharpoons \bar\nu_s$ mixing in the forced adiabatic sweep scenario with lepton number distribution factor 1. However, in Fig.\ \ref{figure7} we see that with large enough $\delta m^2$, negative potential lepton number, and lepton number distribution factor it is possible that the $^4$He yield would fall below $23\%$ by mass fraction, at least for the full resonance sweep scenario.

Similar trends are evident in the deuterium yield as shown in Figures \ref{figure8}, \ref{figure9}, \ref{figure10}, and \ref{figure11}. These figures are essentially similar in overall structure to those for $^4$He. However, because the deuterium yield is so sensitive to baryon-to-photon ratio $\eta$, in these figures we show bands of ranges of calculated D/H corresponding to the quoted uncertainty ranges in $\eta$ for the Three Year WMAP data and for the expected $\eta$ uncertainties in the Four Year WMAP and the Planck results. 

The general change in D/H relative to standard BBN is similar to that for $^4$He. In the lepton number only cases with no sterile neutrinos a positive potential lepton number with its accompanying suppression in $n/p$ results in a decrease in the deuterium abundance yield. Again, this trend is reversed for large enough spectral distortion. In general, bigger increases in D/H are created by larger $\delta m^2$ and larger positive lepton numbers plus $\nu_e \rightleftharpoons \nu_s$ conversion in the full resonance sweep solution.

However, given the current uncertainty in $\eta$ it is evident from these figures that no meaningful constraints on lepton numbers alone or on combinations of lepton numbers and sterile neutrino properties can be obtained from measurements of D/H.  The $\delta$D/H produced, for example, at $\delta m^2 = 10\,{\rm eV}^2$ for $\mathcal{L}_e > 0.15$ could exceed the uncertainty range in deuterium yield stemming from the Three Year WMAP uncertainty in $\eta$, but the $\nu_e$ spectral distortions accompanying this scenario would produce $^4$He in excess of $26\%$ by mass fraction, {\it i.e.}, exceeding the observational bound (horizontal dashed line). 

\begin{figure}
\includegraphics[width=2.5in,angle=270]{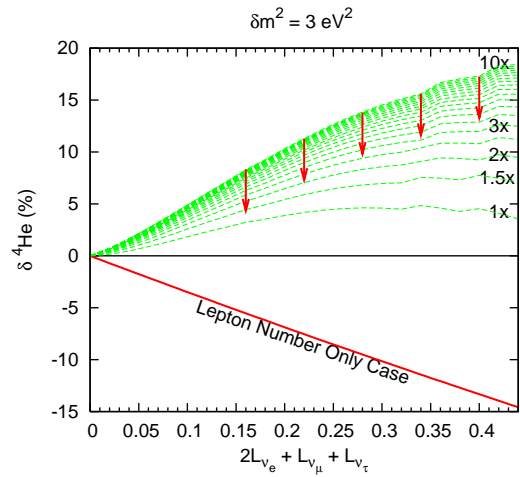}
\caption{Same as for Fig.\ \ref{figure4} but now with $\delta m^2 = 3\,{\rm eV}^2$ and where the dashed curves are for lepton number distribution factors as labeled, up to 10. Here the full resonance sweep solution is employed.}
\label{figure12}
\end{figure}

\begin{figure}
\includegraphics[width=2.5in,angle=270]{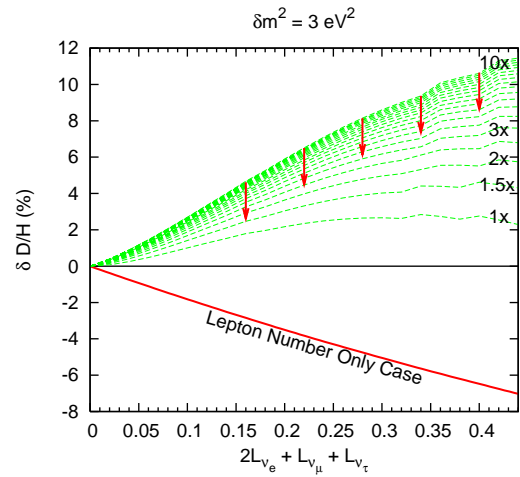}
\caption{Same as Fig.\ \ref{figure12} but now for deuterium.}
\label{figure13}
\end{figure}

By contrast, the considerably smaller uncertainty expected in, {\it e.g.}, the Planck CMB-determined baryon-to-photon ratio could allow for rather stringent constraints on (or signatures of) either lepton numbers alone or combinations of lepton numbers and sterile and active neutrino mixing parameters. However, as is evident in Fig.\ \ref{figure9}, realizing statistically significant constraints or signatures would require that the observationally-derived primordial deuterium abundance be known to better than $5\%$ accuracy. This is problematic as currently there is likely a $15\%$ to $30\%$ error in observationally-determined primordial D/H. Prospects for bettering these errors will be discussed in the next section.

Negative lepton numbers produce alterations in D/H yield which are qualitatively similar (with reversed trends) to those in the positive lepton number regime. However, as for $^4$He, the presence of the threshold in $\bar\nu_e + p \rightarrow n + e^+$ serves to lessen the overall quantitative impact on $\mid\delta {\rm D/H}\mid$ of spectral distortions from given values of (negative) $\mathcal{L}_e$, $\delta m^2$ and lepton number distribution factor. This is shown in Figures \ref{figure10} and \ref{figure11}.

The light element abundance yield alterations resulting from lepton numbers and spectral distortions can depend on $\delta m^2$, $L_{\nu_e}$, $L_{\nu_\mu}$, $L_{\nu_\tau}$, and resonance sweep physics in complicated ways. In Figures \ref{figure12} and \ref{figure13} we show $\delta ^4$He and $\delta$D/H, respectively, for the full resonance sweep solution and for $\delta m^2 = 3\,{\rm eV}^2$ for many values of the lepton number distribution factor $(L_{\nu_{\mu}} + L_{\nu_{\tau}})/{2 L_{\nu_{e}}} =$ 1, 1.5, 2, 2.5, 3, up to 10. We see that there is a fair increase in $\delta ^4$He and $\delta$D/H with increasing values of this factor until it approaches $\approx 5$. In broad brush, this trend comes about because larger values of this factor mean relatively lower $\nu_e$ degeneracy and resonance sweep to higher $\nu_e$ energy ({\it e.g.}, larger $\epsilon_{\rm max}$). Both of these consequences tend to increase $^4$He and D/H yields.

\begin{figure}
\includegraphics[width=2.5in,angle=270]{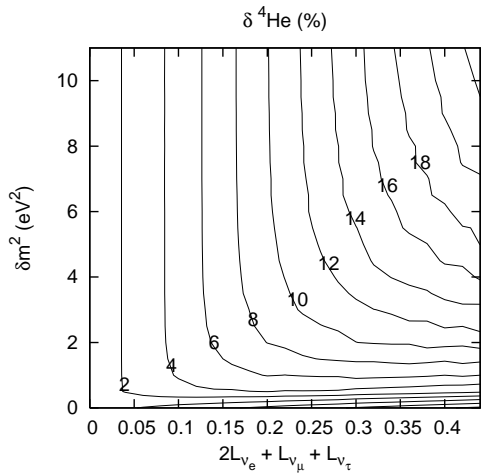}
\caption{Contours of percent change in $^4$He yield relative to standard BBN for ranges of $\delta m^2 = 0 - 11 \,{\rm eV}^2$ and potential lepton number $0 - 0.44$, all for lepton number distribution factor 3 and the full resonance sweep scenario.}
\label{figure14}
\end{figure}

\begin{figure}
\includegraphics[width=2.5in,angle=270]{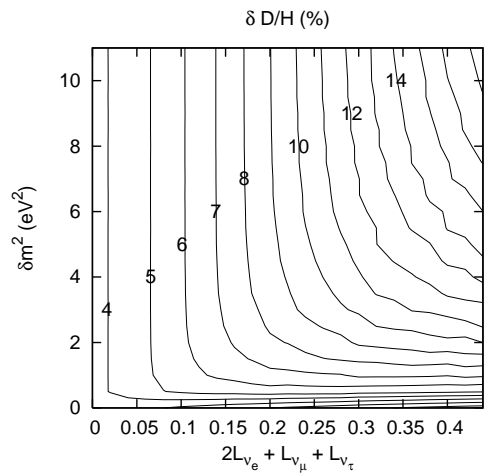}
\caption{Same as Fig.\ \ref{figure14} but now for the primordial deuterium abundance yield.}
\label{figure15}
\end{figure}

Likewise, larger $\delta m^2$ values generally imply an earlier onset of resonance sweep and spectral distortion development. This, in turn, means a bigger effect on $n/p$, as the rates for the neutron-to-proton interconversion processes in Eqs.\ (\ref{nuen}), (\ref{nuebarp}), and (\ref{ndecay}) are faster at earlier epochs where temperature and, hence, average lepton energies are higher. These trends are evident in Figures \ref{figure14} and \ref{figure15} where contours of $\delta ^4$He and $\delta$D/H, respectively, are shown as functions of $\delta m^2$ and (positive) $\mathcal{L}_e = 2L_{\nu_e} + L_{\nu_\mu} + L_{\nu_\tau}$ for the full resonance sweep solution and for lepton number distribution factor 3. There is little dependence of abundance yield deviation on $\delta m^2$ in either figure at low values of $\mathcal{L}_e$. However, for a given $\delta m ^2$, increasing $\mathcal{L}_e$ tends to delay resonance sweep and the development of spectral distortions. This can be offset with larger $\delta m^2$. As a consequence, for larger values of $\mathcal{L}_e$ we see a significant $\delta m^2$ dependence in both $\delta ^4$He and $\delta$D/H.

The results shown in Figures \ref{figure14} and \ref{figure15} may help to indicate where we might expect active neutrino inelastic scattering to partially erase or modify the spectral distortions we calculate in the coherent neutrino propagation limit. The earlier the onset of resonance sweep, the more significant inelastic neutrino scattering will be. This regime will generally be where the $\delta m^2$ dependence in abundance deviations is weakest, ${\it i.e.}$, for parameters in the upper left hand corners of Figures \ref{figure14} and \ref{figure15}, where lepton numbers are small and $\delta m^2$ is large.

\begin{figure}
\includegraphics[width=2.5in,angle=270]{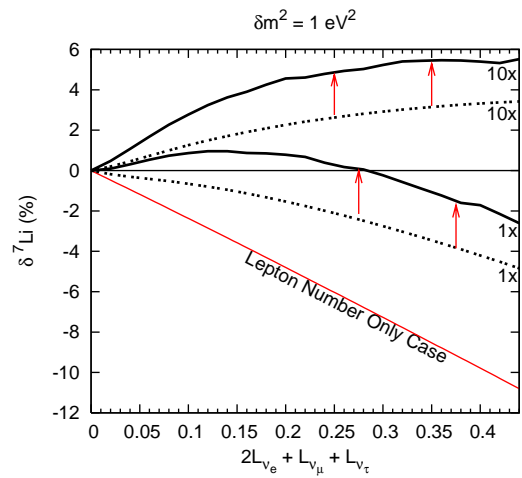}
\caption{Same as Fig.\ \ref{figure4} but now for percent change in $^7$Li abundance.}
\label{figure16}
\end{figure}

\begin{figure}
\includegraphics[width=2.5in,angle=270]{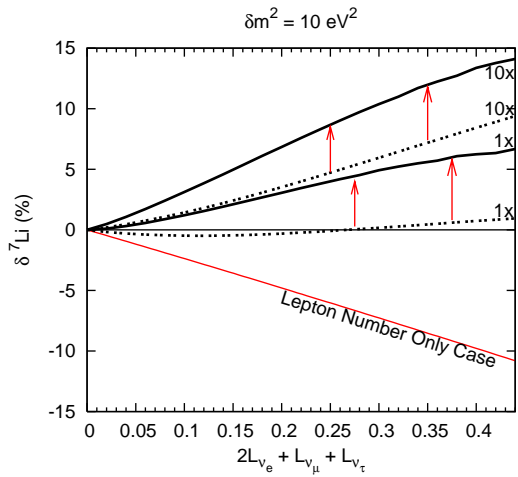}
\caption{Same as Fig.\ \ref{figure16} but now for $\delta m^2 = 10 \,{\rm eV}^2$.}
\label{figure17}
\end{figure}

Finally, the deviation in $^7$Li abundance ($^7$Li/H) yield, $\delta ^7$Li, relative to the standard BBN zero lepton number, no new neutrino physics case, is shown as a function of (positive) potential lepton number $\mathcal{L}_e$ in Figures \ref{figure16} ($\delta m^2 = 1\,{\rm eV}^2$) and \ref{figure17} ($\delta m^2 = 10\,{\rm eV}^2$) for several cases. The general trends in $\delta ^7$Li are similar to those for $\delta ^4$He. In the lepton number only case with no sterile neutrinos, increasing $\mathcal{L}_e$ and, hence, decreasing $n/p$ suppresses the $^7$Li yield. $^7$Li is produced at this $\eta$ principally as $^7$Be through $^3{\rm He}(\alpha, \gamma)^7{\rm Be}$. However, there is a small contribution to $^7$Li from direct production via $^3{\rm H}(\alpha, \gamma)^7{\rm Li}$ and the tritium, $^3$H, abundance during BBN tracks the $n/p$ ratio. Though $\delta ^7$Li can be large for several of the cases shown in these figures, it is at this time not a good candidate for lepton number or sterile neutrino constraint. This is because there remains considerable controversy surrounding both the observationally-determined primordial $^7$Li abundance and the astration/production history of $^7$Li in stars and in the interstellar medium. At the value of the baryon-to-photon ratio $\eta$ adopted here, the calculated BBN $^7$Li/H yield is a factor $\sim 3$ higher than the $^7$Li/H abundance inferred from the Spite plateau in hot, old halo stars \cite{spite}.

\section{Discussion and Conclusions}

We have for the first time self consistently and simultaneously coupled the full BBN nuclear reaction network with medium-enhanced active-sterile neutrino flavor transformation. One conclusion from our work is straightforward: the existence of light sterile neutrinos $\nu_s$ which mix with active neutrinos $\nu_e$, $\nu_\mu$, $\nu_\tau$ could alter significantly the relationship between primordial lepton numbers and the BBN light element ($^2$H, $^4$He, $^7$Li) abundance yields. Our work also shows that precision predictions of primordial nucleosynthesis yields for given neutrino properties and lepton numbers likely will require accurate treatment of the evolution of neutrino spectral distortion. Depending on neutrino mixing parameters, this may require a full $4\times 4$ quantum kinetic equation neutrino flavor transformation scheme.

However, the calculations we have performed here allow us to point out some intriguing trends that eventually may provide a means for constraining lepton numbers and/or sterile neutrino properties. In particular, though we find that positive (negative) potential lepton number causes the BBN $^4$He, $^2$H, and $^7$Li abundance yields to be suppressed (increased) relative to the standard BBN zero-lepton-number case, these trends are counteracted and even {\it reversed} when sterile neutrinos exist and medium-enhanced $\nu_e\rightleftharpoons \nu_s$ ($\bar\nu_e \rightleftharpoons \bar\nu_s$) takes place.

An underlying theme of our work is that the BBN paradigm, especially as it concerns constraints on neutrino physics, may be changing. We now know to fair precision the baryon density from the CMB acoustic peak amplitude ratios. The uncertainty in baryon-to-photon ratio $\eta$ likely will improve with future observations. This trend will culminate in the near term in the Planck mission, which is projected to reduce the uncertainty in $\eta$ to $\leq 1\%$. The near elimination of uncertainty in this quantity and the neutron lifetime will leave the leptonic sector as the principle source of uncertainty in conventional, thermal BBN. This is an intriguing development which comes hard on the heels of new experimental/observational revelations of neutrino mass-squared differences, flavor mixing parameters, and CMB- and large scale structure- derived bounds on light neutrino mass contributions to closure.

The active neutrino mixing parameters in particular allow $^4$He-derived constraints on electron lepton number $L_{\nu_e}$ \cite{Kneller:2001cd} to be extended to $L_{\nu_\mu}$ and $L_{\nu_\tau}$ \cite{abb,wong,dolgov}. Certainly, improvements in precision in the observationally-inferred helium abundance $Y_p$ will translate directly into improved constraints on lepton number \cite{Kneller:2001cd}. This is obvious in Figures \ref{figure4}, \ref{figure5}, \ref{figure6}, \ref{figure7}, for the value of $\mathcal{L}_e = 4L_{\nu_e}$ at which the \lq\lq Lepton Number Only Case\rq\rq\ line crosses the limits of the horizontal band ($^4$He mass fraction $23\%$). This occurs for $L_{\nu_e} = \mathcal{L}_e / 4 \approx 0.045$, corresponding to $\nu_e$ degeneracy parameter $\eta _{\nu_e} \approx 0.07$. This is, of course, a crude upper limit, corresponding to our adopted range for the primordial helium abundance.

If the uncertainty in the observationally-inferred deuterium abundance D/H could be improved significantly, it could become competitive with $^4$He as a probe of lepton number and new neutrino physics. This is not the case currently. The statistical uncertainty in D/H is between $15\%$ and $30\%$, as derived from the isotope-shifted hydrogen absorption lines observed in Lyman Limit and Damped Lyman Alpha absorption systems along lines of sight to high redshift QSO's \cite{Tytler,pettini}. As Figures \ref{figure8}, \ref{figure9}, \ref{figure10}, and \ref{figure11} show, we would need to get the error in D/H down to $\leq 5\%$ to enable deuterium to provide constraints on, or signatures of, {\it e.g.}, sterile neutrinos.

This is problematic in the short term, but not an inconceivable eventuality in the longer term. The current deuterium abundance stems from of order a half dozen QSO absorption systems. With the projected increase in the number of 8m class telescopes and extensions of surveys to the southern hemisphere, we might expect to increase the number of ``clean" ({\it i.e.}, no interloper cloud) Lyman-$\alpha$ absorption systems by a factor of 2 or so. This likely will not be good enough. However, the advent of 30m class telescopes could give many more QSO lines of sight and this might provide for much higher precision deuterium abundance determinations. This could be useful for probes of primordial baryon inhomogeneity or the star formation and chemical evolution history of the early universe. However, in this paper we point out that it also could be useful for constraints on new neutrino physics, especially as regards lepton numbers plus sterile neutrinos.

If, for example, we knew {\it a priori} the sign of the primordial lepton number, then there may be a signature for sterile neutrinos or at least a means for better constraining their properties. A positive lepton number would be expected to give a suppression of $^2$H and $^4$He relative to the zero-lepton-number, standard BBN predictions based on the CMB-derived baryon density. However, a light sterile neutrino could reverse this and give an {\it increase} in these abundances over the standard case. In broad brush, it works in an opposite sense for a negative lepton number, though the effect is less dramatic on account of the threshold in $\bar\nu_e+p\rightarrow n+e^+$. 

Likewise, if experimental evidence exists for active-sterile neutrino mixing, then the sign of the deviation of $^2$H, $^4$He, or $^7$Li from the standard zero-lepton number BBN case may afford a direct measure of the sign and magnitude of the potential lepton number. At the least, measurements and reasoning along these lines may allow for significantly better constraints on lepton numbers in this case.

There is another possibility.  Suppose that active-sterile mixing properties are measured in the lab and precise primordial light-element abundances are obtained from observation.  Further suppose that the effects (abundance deviations from standard BBN) pointed out here are {\it not} seen.  This could indicate that there is another mechanism, other than large ($\vert L_{\nu_\alpha} \vert > 10^{-3}$) lepton number \cite{abfw,FV,cirelli}, operating to suppress active neutrino scattering-induced de-coherence production \cite{vol00,lee00,dol81,mck94,dib00,fv97} of seas of $\nu_s$ and $\bar\nu_s$.  We know that some mechanism must suppress this process because otherwise there would be a conflict with CMB- and large scale structure-derived bounds on light neutrino mass contribution to closure (see for example Ref.\ \cite{abfw}).  Two alternative means for sterile neutrino production suppression have been suggested:  a low re-heat temperature for inflation \cite{gel04}; and an alteration of neutrino mass/mixing properties ({\it i.e.}, no mass or mixing) at early epochs. 

In any case, the linkage between the light elements and new neutrino physics which we have pointed out here increases the leverage of new developments in both observational cosmology and terrestrial neutrino oscillation experiments.

\begin{acknowledgments}
We would like to acknowledge discussions with Huaiyu Duan, David Kirkman, Max Pettini, Michael Smith, Nao Suzuki, and David Tytler.  The work of G.M.F., C.T.K., and C.J.S. was supported in part by NSF grant PHY-04-00359 and a UC/LANL CARE grant at UCSD; C.T.K. acknowledges financial support from the ARCS Foundation, Inc.; K.N.A. was supported by Los Alamos National Laboratory (under DOE contract W-7405-ENG-36) and also acknowledges the UC/LANL CARE grant.
\end{acknowledgments}

\bibliography{DeltaD}

\begin{thebibliography}{54}
\expandafter\ifx\csname natexlab\endcsname\relax\def\natexlab#1{#1}\fi
\expandafter\ifx\csname bibnamefont\endcsname\relax
  \def\bibnamefont#1{#1}\fi
\expandafter\ifx\csname bibfnamefont\endcsname\relax
  \def\bibfnamefont#1{#1}\fi
\expandafter\ifx\csname citenamefont\endcsname\relax
  \def\citenamefont#1{#1}\fi
\expandafter\ifx\csname url\endcsname\relax
  \def\url#1{\texttt{#1}}\fi
\expandafter\ifx\csname urlprefix\endcsname\relax\def\urlprefix{URL }\fi
\providecommand{\bibinfo}[2]{#2}
\providecommand{\eprint}[2][]{\url{#2}}

\bibitem[{\citenamefont{Tegmark et~al.}(2004)}]{WMAP}
\bibinfo{author}{\bibfnamefont{M.}~\bibnamefont{Tegmark}} \bibnamefont{et~al.}
  (\bibinfo{collaboration}{SDSS Collaboration}), \bibinfo{journal}{Phys.\ Rev.\
  D} \textbf{\bibinfo{volume}{69}}, \bibinfo{pages}{103501}
  (\bibinfo{year}{2004}).

\bibitem[{\citenamefont{Spergel et~al.}(2003)}]{WMAP1}
\bibinfo{author}{\bibfnamefont{D.~N.} \bibnamefont{Spergel}}
  \bibnamefont{et~al.}, \bibinfo{journal}{Astrophys.\ J.\ Suppl.}
  \textbf{\bibinfo{volume}{148}}, \bibinfo{pages}{175} (\bibinfo{year}{2003}).

\bibitem[{\citenamefont{Spergel et~al.}(2006)}]{3yrwmap}
\bibinfo{author}{\bibfnamefont{D.~N.} \bibnamefont{Spergel}}
  \bibnamefont{et~al.} (\bibinfo{year}{2006}), \eprint{astro-ph/0603449}.

\bibitem[{\citenamefont{Kirkman et~al.}(2003)\citenamefont{Kirkman, Tytler,
  Suzuki, O'Meara, , and Lubin}}]{Tytler}
\bibinfo{author}{\bibfnamefont{D.}~\bibnamefont{Kirkman}},
  \bibinfo{author}{\bibfnamefont{D.}~\bibnamefont{Tytler}},
  \bibinfo{author}{\bibfnamefont{N.}~\bibnamefont{Suzuki}},
  \bibinfo{author}{\bibfnamefont{J.~M.} \bibnamefont{O'Meara}}, ,
  \bibnamefont{and} \bibinfo{author}{\bibfnamefont{D.}~\bibnamefont{Lubin}},
  \bibinfo{journal}{Astrophys.\ J.\ Suppl.} \textbf{\bibinfo{volume}{149}},
  \bibinfo{pages}{1} (\bibinfo{year}{2003}).

\bibitem[{\citenamefont{O'Meara et~al.}(2001)\citenamefont{O'Meara, Tytler,
  Kirkman, Suzuki, Prochaska, Lubin, and Wolfe}}]{OMeara}
\bibinfo{author}{\bibfnamefont{J.~M.} \bibnamefont{O'Meara}},
  \bibinfo{author}{\bibfnamefont{D.}~\bibnamefont{Tytler}},
  \bibinfo{author}{\bibfnamefont{D.}~\bibnamefont{Kirkman}},
  \bibinfo{author}{\bibfnamefont{N.}~\bibnamefont{Suzuki}},
  \bibinfo{author}{\bibfnamefont{J.~X.} \bibnamefont{Prochaska}},
  \bibinfo{author}{\bibfnamefont{D.}~\bibnamefont{Lubin}}, \bibnamefont{and}
  \bibinfo{author}{\bibfnamefont{A.~M.} \bibnamefont{Wolfe}},
  \bibinfo{journal}{Astrophys.\ J.} \textbf{\bibinfo{volume}{552}},
  \bibinfo{pages}{718} (\bibinfo{year}{2001}).

\bibitem[{\citenamefont{Burles et~al.}(2001)\citenamefont{Burles, Nollett, and
  Turner}}]{burles}
\bibinfo{author}{\bibfnamefont{S.}~\bibnamefont{Burles}},
  \bibinfo{author}{\bibfnamefont{K.~M.} \bibnamefont{Nollett}},
  \bibnamefont{and} \bibinfo{author}{\bibfnamefont{M.~S.}
  \bibnamefont{Turner}}, \bibinfo{journal}{Astrophys.\ J.}
  \textbf{\bibinfo{volume}{552}}, \bibinfo{pages}{L1} (\bibinfo{year}{2001}).

\bibitem[{\citenamefont{Cyburt et~al.}(2003)\citenamefont{Cyburt, Fields, and
  Olive}}]{cyb03}
\bibinfo{author}{\bibfnamefont{R.~H.} \bibnamefont{Cyburt}},
  \bibinfo{author}{\bibfnamefont{B.~D.} \bibnamefont{Fields}},
  \bibnamefont{and} \bibinfo{author}{\bibfnamefont{K.~A.} \bibnamefont{Olive}},
  \bibinfo{journal}{Phys.\ Lett.\ B} \textbf{\bibinfo{volume}{567}},
  \bibinfo{pages}{227} (\bibinfo{year}{2003}).

\bibitem[{\citenamefont{Cardall and Fuller.}(1996)}]{CF}
\bibinfo{author}{\bibfnamefont{C.~Y.} \bibnamefont{Cardall}} \bibnamefont{and}
  \bibinfo{author}{\bibfnamefont{G.~M.} \bibnamefont{Fuller.}},
  \bibinfo{journal}{Astrophys.\ J.} \textbf{\bibinfo{volume}{472}},
  \bibinfo{pages}{435} (\bibinfo{year}{1996}).

\bibitem[{\citenamefont{Abazajian et~al.}(2005)\citenamefont{Abazajian, Bell,
  Fuller, and Wong}}]{abfw}
\bibinfo{author}{\bibfnamefont{K.}~\bibnamefont{Abazajian}},
  \bibinfo{author}{\bibfnamefont{N.~F.} \bibnamefont{Bell}},
  \bibinfo{author}{\bibfnamefont{G.~M.} \bibnamefont{Fuller}},
  \bibnamefont{and} \bibinfo{author}{\bibfnamefont{Y.~Y.~Y.}
  \bibnamefont{Wong}}, \bibinfo{journal}{Phys.\ Rev.\ D}
  \textbf{\bibinfo{volume}{72}}, \bibinfo{pages}{063004}
  (\bibinfo{year}{2005}).

\bibitem[{\citenamefont{Kishimoto et~al.}(2006)\citenamefont{Kishimoto, Fuller,
  and Smith}}]{kfs}
\bibinfo{author}{\bibfnamefont{C.~T.} \bibnamefont{Kishimoto}},
  \bibinfo{author}{\bibfnamefont{G.~M.} \bibnamefont{Fuller}},
  \bibnamefont{and} \bibinfo{author}{\bibfnamefont{C.~J.} \bibnamefont{Smith}}
  (\bibinfo{year}{2006}), \eprint{astro-ph/0607403}.

\bibitem[{\citenamefont{Olive and Skillman}(2004)}]{OS}
\bibinfo{author}{\bibfnamefont{K.~A.} \bibnamefont{Olive}} \bibnamefont{and}
  \bibinfo{author}{\bibfnamefont{E.~D.} \bibnamefont{Skillman}},
  \bibinfo{journal}{Astrophys.\ J.} \textbf{\bibinfo{volume}{617}},
  \bibinfo{pages}{29} (\bibinfo{year}{2004}).

\bibitem[{\citenamefont{Olive et~al.}(1997)\citenamefont{Olive, Steigman, and
  Skillman}}]{Olive}
\bibinfo{author}{\bibfnamefont{K.~A.} \bibnamefont{Olive}},
  \bibinfo{author}{\bibfnamefont{G.}~\bibnamefont{Steigman}}, \bibnamefont{and}
  \bibinfo{author}{\bibfnamefont{E.~D.} \bibnamefont{Skillman}},
  \bibinfo{journal}{Astrophys.\ J.} \textbf{\bibinfo{volume}{483}},
  \bibinfo{pages}{788} (\bibinfo{year}{1997}).

\bibitem[{\citenamefont{Izotov and Thuan}(2004)}]{IT}
\bibinfo{author}{\bibfnamefont{Y.~I.} \bibnamefont{Izotov}} \bibnamefont{and}
  \bibinfo{author}{\bibfnamefont{T.~X.} \bibnamefont{Thuan}},
  \bibinfo{journal}{Astrophys.\ J.} \textbf{\bibinfo{volume}{602}},
  \bibinfo{pages}{200} (\bibinfo{year}{2004}).

\bibitem[{\citenamefont{Cyburt et~al.}(2005)\citenamefont{Cyburt, Fields,
  Olive, and Skillman}}]{cyburt}
\bibinfo{author}{\bibfnamefont{R.~H.} \bibnamefont{Cyburt}},
  \bibinfo{author}{\bibfnamefont{B.~D.} \bibnamefont{Fields}},
  \bibinfo{author}{\bibfnamefont{K.~A.} \bibnamefont{Olive}}, \bibnamefont{and}
  \bibinfo{author}{\bibfnamefont{E.}~\bibnamefont{Skillman}},
  \bibinfo{journal}{Astropart.\ Phys.} \textbf{\bibinfo{volume}{23}},
  \bibinfo{pages}{313} (\bibinfo{year}{2005}).

\bibitem[{\citenamefont{Steigman}(2005)}]{Steigman}
\bibinfo{author}{\bibfnamefont{G.}~\bibnamefont{Steigman}}
  (\bibinfo{year}{2005}), \eprint{astro-ph/0501591}.

\bibitem[{\citenamefont{Steigman}(2006)}]{Steigman1}
\bibinfo{author}{\bibfnamefont{G.}~\bibnamefont{Steigman}},
  \bibinfo{journal}{Int.\ J.\ Mod.\ Phys.\ E} \textbf{\bibinfo{volume}{15}},
  \bibinfo{pages}{1} (\bibinfo{year}{2006}).

\bibitem[{\citenamefont{Pettini}(2006)}]{pettini}
\bibinfo{author}{\bibfnamefont{M.}~\bibnamefont{Pettini}}, in
  \emph{\bibinfo{booktitle}{Astrophysics in the Far Ultraviolet: Five Years of
  Discovery with FUSE}}, edited by
  \bibinfo{editor}{\bibfnamefont{G.}~\bibnamefont{Sonneborn}},
  \bibinfo{editor}{\bibfnamefont{H.~W.} \bibnamefont{Moos}}, \bibnamefont{and}
  \bibinfo{editor}{\bibfnamefont{B.-G.} \bibnamefont{Andersson}}
  (\bibinfo{publisher}{ASP}, \bibinfo{address}{Provo}, \bibinfo{year}{2006}),
  ASP Conf.\ Ser.\ No. 348, p.~\bibinfo{pages}{19}.

\bibitem[{\citenamefont{Eitel}(2000)}]{LSND}
\bibinfo{author}{\bibfnamefont{K.}~\bibnamefont{Eitel}}, \bibinfo{journal}{New
  J.\ Phys.} \textbf{\bibinfo{volume}{2}}, \bibinfo{pages}{1}
  (\bibinfo{year}{2000}).

\bibitem[{\citenamefont{McGregor}(2003)}]{miniB}
\bibinfo{author}{\bibfnamefont{G.}~\bibnamefont{McGregor}}
  (\bibinfo{collaboration}{MiniBooNE Collaboration}), in
  \emph{\bibinfo{booktitle}{Particle Physics and Cosmology: Third Tropical
  Workshop on Particle Physics and Cosmology - Neutrinos, Branes, and
  Cosmology}}, edited by \bibinfo{editor}{\bibfnamefont{J.~F.}
  \bibnamefont{Nieves}} \bibnamefont{and} \bibinfo{editor}{\bibfnamefont{C.~N.}
  \bibnamefont{Leung}} (\bibinfo{publisher}{AIP}, \bibinfo{address}{New York},
  \bibinfo{year}{2003}), AIP Conf.\ Proc.\ No.\ 655, p.~\bibinfo{pages}{58}.

\bibitem[{\citenamefont{Seljak et~al.}(2006)\citenamefont{Seljak, Slosar, and
  McDonald}}]{Sloan}
\bibinfo{author}{\bibfnamefont{U.}~\bibnamefont{Seljak}},
  \bibinfo{author}{\bibfnamefont{A.}~\bibnamefont{Slosar}}, \bibnamefont{and}
  \bibinfo{author}{\bibfnamefont{P.}~\bibnamefont{McDonald}}
  (\bibinfo{year}{2006}), \eprint{astro-ph/0604335}.

\bibitem[{\citenamefont{Pierce and Murayama}(2004)}]{murayama}
\bibinfo{author}{\bibfnamefont{A.}~\bibnamefont{Pierce}} \bibnamefont{and}
  \bibinfo{author}{\bibfnamefont{H.}~\bibnamefont{Murayama}},
  \bibinfo{journal}{Phys.\ Lett.\ B} \textbf{\bibinfo{volume}{581}},
  \bibinfo{pages}{218} (\bibinfo{year}{2004}).

\bibitem[{\citenamefont{Abazajian}(2003)}]{aba}
\bibinfo{author}{\bibfnamefont{K.~N.} \bibnamefont{Abazajian}},
  \bibinfo{journal}{Astropart.\ Phys.} \textbf{\bibinfo{volume}{19}},
  \bibinfo{pages}{303} (\bibinfo{year}{2003}).

\bibitem[{\citenamefont{Hannestad}(2003)}]{hannestad}
\bibinfo{author}{\bibfnamefont{S.}~\bibnamefont{Hannestad}},
  \bibinfo{journal}{J.\ Cosmol.\ Astropart.\ Phys}
  \textbf{\bibinfo{volume}{0305}}, \bibinfo{pages}{004} (\bibinfo{year}{2003}).

\bibitem[{\citenamefont{{Di Bari}}(2002)}]{dib}
\bibinfo{author}{\bibfnamefont{P.}~\bibnamefont{{Di Bari}}},
  \bibinfo{journal}{Phys.\ Rev.\ D} \textbf{\bibinfo{volume}{65}},
  \bibinfo{pages}{043509} (\bibinfo{year}{2002}).

\bibitem[{\citenamefont{{Di Bari}}(2003)}]{dibad}
\bibinfo{author}{\bibfnamefont{P.}~\bibnamefont{{Di Bari}}},
  \bibinfo{journal}{Phys.\ Rev.\ D} \textbf{\bibinfo{volume}{67}},
  \bibinfo{pages}{127301} (\bibinfo{year}{2003}).

\bibitem[{\citenamefont{Barger et~al.}(2003)\citenamefont{Barger, Kneller,
  Langacker, Marfatia, and Steigman}}]{barger}
\bibinfo{author}{\bibfnamefont{V.}~\bibnamefont{Barger}},
  \bibinfo{author}{\bibfnamefont{J.~P.} \bibnamefont{Kneller}},
  \bibinfo{author}{\bibfnamefont{P.}~\bibnamefont{Langacker}},
  \bibinfo{author}{\bibfnamefont{D.}~\bibnamefont{Marfatia}}, \bibnamefont{and}
  \bibinfo{author}{\bibfnamefont{G.}~\bibnamefont{Steigman}},
  \bibinfo{journal}{Phys.\ Lett.\ B} \textbf{\bibinfo{volume}{569}},
  \bibinfo{pages}{123} (\bibinfo{year}{2003}).

\bibitem[{\citenamefont{Kneller et~al.}(2001)\citenamefont{Kneller, Scherrer,
  Steigman, and Walker}}]{Kneller:2001cd}
\bibinfo{author}{\bibfnamefont{J.~P.} \bibnamefont{Kneller}},
  \bibinfo{author}{\bibfnamefont{R.~J.} \bibnamefont{Scherrer}},
  \bibinfo{author}{\bibfnamefont{G.}~\bibnamefont{Steigman}}, \bibnamefont{and}
  \bibinfo{author}{\bibfnamefont{T.~P.} \bibnamefont{Walker}},
  \bibinfo{journal}{Phys.\ Rev.\ D} \textbf{\bibinfo{volume}{64}},
  \bibinfo{pages}{123506} (\bibinfo{year}{2001}).

\bibitem[{\citenamefont{Dodelson et~al.}(2006)\citenamefont{Dodelson,
  Melchiorri, and Slosar}}]{dod06}
\bibinfo{author}{\bibfnamefont{S.}~\bibnamefont{Dodelson}},
  \bibinfo{author}{\bibfnamefont{A.}~\bibnamefont{Melchiorri}},
  \bibnamefont{and} \bibinfo{author}{\bibfnamefont{A.}~\bibnamefont{Slosar}},
  \bibinfo{journal}{Phys.\ Rev.\ Lett.} \textbf{\bibinfo{volume}{97}},
  \bibinfo{pages}{041301} (\bibinfo{year}{2006}).

\bibitem[{\citenamefont{Foot and Volkas}(1995)}]{FV}
\bibinfo{author}{\bibfnamefont{R.}~\bibnamefont{Foot}} \bibnamefont{and}
  \bibinfo{author}{\bibfnamefont{R.~R.} \bibnamefont{Volkas}},
  \bibinfo{journal}{Phys.\ Rev.\ Lett.} \textbf{\bibinfo{volume}{75}},
  \bibinfo{pages}{4350} (\bibinfo{year}{1995}).

\bibitem[{\citenamefont{Foot and Volkas}(1997)}]{fv97}
\bibinfo{author}{\bibfnamefont{R.}~\bibnamefont{Foot}} \bibnamefont{and}
  \bibinfo{author}{\bibfnamefont{R.~R.} \bibnamefont{Volkas}},
  \bibinfo{journal}{Phys.\ Rev.\ D} \textbf{\bibinfo{volume}{55}},
  \bibinfo{pages}{5147} (\bibinfo{year}{1997}).

\bibitem[{\citenamefont{Chu and Cirelli}(2006)}]{cirelli}
\bibinfo{author}{\bibfnamefont{Y.-Z.} \bibnamefont{Chu}} \bibnamefont{and}
  \bibinfo{author}{\bibfnamefont{M.}~\bibnamefont{Cirelli}}
  (\bibinfo{year}{2006}), \eprint{astro-ph/0608206}.

\bibitem[{\citenamefont{Smith et~al.}(1993)\citenamefont{Smith, Kawano, and
  Malaney}}]{Smith:1992yy}
\bibinfo{author}{\bibfnamefont{M.~S.} \bibnamefont{Smith}},
  \bibinfo{author}{\bibfnamefont{L.~H.} \bibnamefont{Kawano}},
  \bibnamefont{and} \bibinfo{author}{\bibfnamefont{R.~A.}
  \bibnamefont{Malaney}}, \bibinfo{journal}{Astrophys.\ J.\ Suppl.}
  \textbf{\bibinfo{volume}{85}}, \bibinfo{pages}{219} (\bibinfo{year}{1993}).

\bibitem[{\citenamefont{Shi and Fuller}(1999)}]{SF}
\bibinfo{author}{\bibfnamefont{X.}~\bibnamefont{Shi}} \bibnamefont{and}
  \bibinfo{author}{\bibfnamefont{G.~M.} \bibnamefont{Fuller}},
  \bibinfo{journal}{Phys.\ Rev.\ Lett.} \textbf{\bibinfo{volume}{83}},
  \bibinfo{pages}{3120} (\bibinfo{year}{1999}).

\bibitem[{\citenamefont{Mikheyev and Smirnov}(1985)}]{MS}
\bibinfo{author}{\bibfnamefont{S.~P.} \bibnamefont{Mikheyev}} \bibnamefont{and}
  \bibinfo{author}{\bibfnamefont{A.~Y.} \bibnamefont{Smirnov}},
  \bibinfo{journal}{Yad.\ Fiz.} \textbf{\bibinfo{volume}{42}},
  \bibinfo{pages}{1441} (\bibinfo{year}{1985}).

\bibitem[{\citenamefont{Wolfenstein}(1978)}]{W}
\bibinfo{author}{\bibfnamefont{L.}~\bibnamefont{Wolfenstein}},
  \bibinfo{journal}{Phys.\ Rev.\ D} \textbf{\bibinfo{volume}{17}},
  \bibinfo{pages}{2369} (\bibinfo{year}{1978}).

\bibitem[{\citenamefont{Wagoner et~al.}(1967)\citenamefont{Wagoner, Fowler, and
  Hoyle}}]{wfh}
\bibinfo{author}{\bibfnamefont{R.~V.} \bibnamefont{Wagoner}},
  \bibinfo{author}{\bibfnamefont{W.~A.} \bibnamefont{Fowler}},
  \bibnamefont{and} \bibinfo{author}{\bibfnamefont{F.}~\bibnamefont{Hoyle}},
  \bibinfo{journal}{Astrophys.\ J.} \textbf{\bibinfo{volume}{148}},
  \bibinfo{pages}{3} (\bibinfo{year}{1967}).

\bibitem[{\citenamefont{Kawano}(1992)}]{kawano}
\bibinfo{author}{\bibfnamefont{L.}~\bibnamefont{Kawano}},
  \bibinfo{journal}{NASA STI/Recon Technical Report N}
  \textbf{\bibinfo{volume}{92}}, \bibinfo{pages}{25163} (\bibinfo{year}{1992}).

\bibitem[{\citenamefont{Kawano}(1988)}]{kawano1}
\bibinfo{author}{\bibfnamefont{L.}~\bibnamefont{Kawano}}
  (\bibinfo{year}{1988}), \eprint{FERMILAB-PUB-88/34-A}.

\bibitem[{\citenamefont{Yao et~al.}(2006)}]{pdg}
\bibinfo{author}{\bibfnamefont{W.-M.} \bibnamefont{Yao}} \bibnamefont{et~al.},
  \bibinfo{journal}{J.\ Phys.\ G} \textbf{\bibinfo{volume}{33}},
  \bibinfo{pages}{1} (\bibinfo{year}{2006}).

\bibitem[{\citenamefont{Serebrov et~al.}(2005)}]{serebrov}
\bibinfo{author}{\bibfnamefont{A.}~\bibnamefont{Serebrov}}
  \bibnamefont{et~al.}, \bibinfo{journal}{Phys.\ Lett.\ B}
  \textbf{\bibinfo{volume}{605}}, \bibinfo{pages}{72} (\bibinfo{year}{2005}).

\bibitem[{\citenamefont{Bond et~al.}(2004)\citenamefont{Bond, Contaldi, Lewis,
  and Pogosyan}}]{bond}
\bibinfo{author}{\bibfnamefont{J.~R.} \bibnamefont{Bond}},
  \bibinfo{author}{\bibfnamefont{C.}~\bibnamefont{Contaldi}},
  \bibinfo{author}{\bibfnamefont{A.}~\bibnamefont{Lewis}}, \bibnamefont{and}
  \bibinfo{author}{\bibfnamefont{D.}~\bibnamefont{Pogosyan}},
  \bibinfo{journal}{Int.\ J.\ Theor.\ Phys.} \textbf{\bibinfo{volume}{43}},
  \bibinfo{pages}{599} (\bibinfo{year}{2004}).

\bibitem[{\citenamefont{Lopez and Turner}(1999)}]{tandl}
\bibinfo{author}{\bibfnamefont{R.~E.} \bibnamefont{Lopez}} \bibnamefont{and}
  \bibinfo{author}{\bibfnamefont{M.~S.} \bibnamefont{Turner}},
  \bibinfo{journal}{Phys.\ Rev.\ D} \textbf{\bibinfo{volume}{59}},
  \bibinfo{pages}{103502} (\bibinfo{year}{1999}).

\bibitem[{\citenamefont{Kernan and Krauss}(1994)}]{kandk}
\bibinfo{author}{\bibfnamefont{P.~J.} \bibnamefont{Kernan}} \bibnamefont{and}
  \bibinfo{author}{\bibfnamefont{L.~M.} \bibnamefont{Krauss}},
  \bibinfo{journal}{Phys.\ Rev.\ Lett.} \textbf{\bibinfo{volume}{72}},
  \bibinfo{pages}{3309} (\bibinfo{year}{1994}).

\bibitem[{\citenamefont{Abazajian et~al.}(2002)\citenamefont{Abazajian, Beacom,
  and Bell}}]{abb}
\bibinfo{author}{\bibfnamefont{K.~N.} \bibnamefont{Abazajian}},
  \bibinfo{author}{\bibfnamefont{J.~F.} \bibnamefont{Beacom}},
  \bibnamefont{and} \bibinfo{author}{\bibfnamefont{N.~F.} \bibnamefont{Bell}},
  \bibinfo{journal}{Phys.\ Rev.\ D} \textbf{\bibinfo{volume}{66}},
  \bibinfo{pages}{013008} (\bibinfo{year}{2002}).

\bibitem[{\citenamefont{Wong}(2002)}]{wong}
\bibinfo{author}{\bibfnamefont{Y.~Y.~Y.} \bibnamefont{Wong}},
  \bibinfo{journal}{Phys.\ Rev.\ D} \textbf{\bibinfo{volume}{66}},
  \bibinfo{pages}{025015} (\bibinfo{year}{2002}).

\bibitem[{\citenamefont{Dolgov et~al.}(2002)\citenamefont{Dolgov, Hansen,
  Pastor, Petcov, Raffelt, and Semikoz}}]{dolgov}
\bibinfo{author}{\bibfnamefont{A.~D.} \bibnamefont{Dolgov}},
  \bibinfo{author}{\bibfnamefont{S.~H.} \bibnamefont{Hansen}},
  \bibinfo{author}{\bibfnamefont{S.}~\bibnamefont{Pastor}},
  \bibinfo{author}{\bibfnamefont{S.~T.} \bibnamefont{Petcov}},
  \bibinfo{author}{\bibfnamefont{G.~G.} \bibnamefont{Raffelt}},
  \bibnamefont{and} \bibinfo{author}{\bibfnamefont{D.~V.}
  \bibnamefont{Semikoz}}, \bibinfo{journal}{Nucl.\ Phys.\ B}
  \textbf{\bibinfo{volume}{632}}, \bibinfo{pages}{363} (\bibinfo{year}{2002}).

\bibitem[{\citenamefont{Savage et~al.}(1991)\citenamefont{Savage, Malaney, and
  Fuller}}]{smf}
\bibinfo{author}{\bibfnamefont{M.~J.} \bibnamefont{Savage}},
  \bibinfo{author}{\bibfnamefont{R.~A.} \bibnamefont{Malaney}},
  \bibnamefont{and} \bibinfo{author}{\bibfnamefont{G.~M.}
  \bibnamefont{Fuller}}, \bibinfo{journal}{Astrophys.\ J.}
  \textbf{\bibinfo{volume}{368}}, \bibinfo{pages}{1} (\bibinfo{year}{1991}).

\bibitem[{\citenamefont{Spite and Spite}(1982)}]{spite}
\bibinfo{author}{\bibfnamefont{F.}~\bibnamefont{Spite}} \bibnamefont{and}
  \bibinfo{author}{\bibfnamefont{M.}~\bibnamefont{Spite}},
  \bibinfo{journal}{Astron.\ Astrophys.} \textbf{\bibinfo{volume}{115}},
  \bibinfo{pages}{357} (\bibinfo{year}{1982}).

\bibitem[{\citenamefont{Volkas and Wong}(2000)}]{vol00}
\bibinfo{author}{\bibfnamefont{R.~R.} \bibnamefont{Volkas}} \bibnamefont{and}
  \bibinfo{author}{\bibfnamefont{Y.~Y.~Y.} \bibnamefont{Wong}},
  \bibinfo{journal}{Phys.\ Rev.\ D} \textbf{\bibinfo{volume}{62}},
  \bibinfo{pages}{093024} (\bibinfo{year}{2000}).

\bibitem[{\citenamefont{Lee et~al.}(2000)\citenamefont{Lee, Volkas, and
  Wong}}]{lee00}
\bibinfo{author}{\bibfnamefont{K.~S.~M.} \bibnamefont{Lee}},
  \bibinfo{author}{\bibfnamefont{R.~R.} \bibnamefont{Volkas}},
  \bibnamefont{and} \bibinfo{author}{\bibfnamefont{Y.~Y.~Y.}
  \bibnamefont{Wong}}, \bibinfo{journal}{Phys.\ Rev.\ D}
  \textbf{\bibinfo{volume}{62}}, \bibinfo{pages}{093025}
  (\bibinfo{year}{2000}).

\bibitem[{\citenamefont{Dolgov}(1981)}]{dol81}
\bibinfo{author}{\bibfnamefont{A.~D.} \bibnamefont{Dolgov}},
  \bibinfo{journal}{Yad.\ Fiz.} \textbf{\bibinfo{volume}{33}},
  \bibinfo{pages}{1309} (\bibinfo{year}{1981}).

\bibitem[{\citenamefont{McKellar and Thomson}(1994)}]{mck94}
\bibinfo{author}{\bibfnamefont{B.~H.~J.} \bibnamefont{McKellar}}
  \bibnamefont{and} \bibinfo{author}{\bibfnamefont{M.~J.}
  \bibnamefont{Thomson}}, \bibinfo{journal}{Phys.\ Rev.\ D}
  \textbf{\bibinfo{volume}{49}}, \bibinfo{pages}{2710} (\bibinfo{year}{1994}).

\bibitem[{\citenamefont{{Di Bari} et~al.}(2000)\citenamefont{{Di Bari}, Lipari,
  and Lusignoli}}]{dib00}
\bibinfo{author}{\bibfnamefont{P.}~\bibnamefont{{Di Bari}}},
  \bibinfo{author}{\bibfnamefont{P.}~\bibnamefont{Lipari}}, \bibnamefont{and}
  \bibinfo{author}{\bibfnamefont{M.}~\bibnamefont{Lusignoli}},
  \bibinfo{journal}{Int.\ J.\ Mod.\ Phys.\ A} \textbf{\bibinfo{volume}{15}},
  \bibinfo{pages}{2289} (\bibinfo{year}{2000}).

\bibitem[{\citenamefont{Gelmini et~al.}(2004)\citenamefont{Gelmini,
  Palomares-Ruiz, and Pascoli}}]{gel04}
\bibinfo{author}{\bibfnamefont{G.}~\bibnamefont{Gelmini}},
  \bibinfo{author}{\bibfnamefont{S.}~\bibnamefont{Palomares-Ruiz}},
  \bibnamefont{and} \bibinfo{author}{\bibfnamefont{S.}~\bibnamefont{Pascoli}},
  \bibinfo{journal}{Phys.\ Rev.\ Lett.} \textbf{\bibinfo{volume}{93}},
  \bibinfo{pages}{081302} (\bibinfo{year}{2004}).

\end{thebibliography}

\end{document}